\documentclass[a4paper,11pt]{article}

\usepackage{jheppub} % for details on the use of the package, please
\usepackage{color}
\usepackage{lineno}
\title{Cryogenic light detectors with thermal signal amplification for $0\nu\beta\beta$ search experiments}

\author[a,1]{A.~Armatol,\note{Now at Institut de Physique des 2 Infinis, Villeurbanne, France.}}
\author[b]{A.S.~Barabash,}
\author[a]{D.~Baudin,}
\author[a,2]{V.~Berest,\note{Corresponding author.}}
\author[c,3]{M.~Beretta,\note{Now at University of Milano-Bicocca and INFN Sezione di Milano-Bicocca, Milan, Italy.}}
\author[d]{L.~Berg{\'e},}
\author[d,4]{M.~Buchynska,\note{Corresponding author.}}
\author[e,f]{J.M.~Calvo-Mozota,}
\author[g,5]{C.~Capelli,\note{Now at at Physik-Institut, University of Z\"urich, Z\"urich,
Switzerland.}}
\author[h,i]{P.~Carniti,}
\author[d]{M.~Chapellier,}
\author[j]{I.~Dafinei,}
\author[k,l]{F.A.~Danevich,}
\author[a,d,6]{T.~Dixon,\note{Now at University College London, London, UK.}}
\author[g]{A.~Drobizhev,}
\author[d]{L.~Dumoulin,}
\author[a]{F.~Ferri,}
\author[d]{A.~Gallas,}
\author[d]{A.~Giuliani,}
\author[i,h]{C.~Gotti,}
\author[a]{Ph.~Gras,}
\author[m]{A.~Ianni,}
\author[d,7]{L.~Imbert,\note{Now at INFN Sezione di Milano-Bicocca, Milan, Italy.}}
\author[a]{H.~Khalife,}
\author[k]{V.V.~Kobychev,}
\author[b]{S.I.~Konovalov,}
\author[d]{P.~Loaiza,}
\author[d]{P.~de~Marcillac,}
\author[d]{S.~Marnieros,}
\author[d]{C.A.~Marrache-Kikuchi,}
\author[n,o]{M.~Martinez,}
\author[a]{E.~Mazzucato,}
\author[a]{C.~Nones,}
\author[d]{E.~Olivieri,}
\author[n]{A.~Ortiz de Sol\'orzano,}
\author[a]{M.~Pageot,}
\author[d]{Y.~Peinaud,}
\author[d,e]{V.~P\'erez,}
\author[i,h]{G.~Pessina,}
\author[d]{D.V.~Poda,}
\author[d]{P.~Rosier,}
\author[d]{J.A.~Scarpaci,}
\author[a]{B.~Schmidt,}
\author[k,l]{V.I.~Tretyak,}
\author[b]{V.I.~Umatov,}
\author[k]{M.M.~Zarytskyy,}
\author[a]{A.~Zolotarova}

\affiliation[a]{IRFU, CEA, Université Paris-Saclay, 91191 Saclay, France}
\affiliation[b]{National Research Center Kurchatov Institute, Kurchatov Complex of Theoretical and Experimental Physics, 117218 Moscow, Russia}
\affiliation[c]{University of California, Berkeley-94720, CA, USA}
\affiliation[d]{Universit\'e Paris-Saclay, CNRS/IN2P3, IJCLab, 91405 Orsay, France}
\affiliation[e]{Laboratorio Subterr\'aneo de Canfranc, 22880 Canfranc-Estaci\'on, Spain}
\affiliation[f]{Escuela Superior de Ingenier\'ia y Tecnolog\'ia, Universidad Internacional de La Rioja, 26006 Logro\~no, Spain}
\affiliation[g]{Lawrence Berkeley National Laboratory, Berkeley, CA 94720, USA}
\affiliation[h]{INFN Sezione di Milano-Bicocca, I-20126 Milan, Italy}
\affiliation[i]{University of Milano-Bicocca, I-20126 Milan, Italy}
\affiliation[j]{INFN Sezione di Roma, I-00185 Rome, Italy}
\affiliation[k]{Institute for Nuclear Research of NASU, 03028 Kyiv, Ukraine}
\affiliation[l]{Institute of Experimental and Applied Physics, CTU Prague, Prague, CA-11000, Czech Republic}
\affiliation[m]{INFN Laboratori Nazionali del Gran Sasso, I-67100 Assergi (AQ), Italy}
\affiliation[n]{Centro de Astropart\'iculas y F\'isica de Altas Energ\'ias, Universidad de Zaragoza, 50009 Zaragoza, Spain}
\affiliation[o]{ARAID Fundaci\'on Agencia Aragonesa para la Investigaci\'on y el Desarrollo, 50018 Zaragoza, Spain}

\emailAdd{vladyslav.berest@cea.fr}
\emailAdd{mariia.buchynska@ijclab.in2p3.fr}

\abstract{
As a step towards the realization of cryogenic-detector experiments to search for neutrinoless double-beta decay (such as CROSS, BINGO, and CUPID), we investigated a batch of 10 Ge light detectors (LDs) assisted by Neganov-Trofimov-Luke (NTL) signal amplification. Each LD was assembled with a large cubic light-emitting crystal (45 mm side) using the recently developed CROSS mechanical structure. The detector array was operated at milli-Kelvin temperatures in a pulse-tube cryostat at the Canfranc underground laboratory in Spain. We achieved good performance with scintillating bolometers from CROSS, made of Li$_{2}${}$^{100}$MoO$_4$ crystals and used as reference detectors of the setup, and with all LDs tested (except for a single device that encountered an electronics issue). No leakage current was observed for 8 LDs with an electrode bias up to 100 V. Operating the LDs at an 80 V electrode bias applied in parallel, we obtained a gain of around 9 in the signal-to-noise ratio of these devices, allowing us to achieve a baseline noise RMS of $O$(10~eV). Thanks to the strong current polarization of the temperature sensors, the time response of the devices was reduced to around half a millisecond in rise time. The achieved performance of the LDs was extrapolated via simulations of pile-up rejection capability for several configurations of the CUPID detector structure. Despite the sub-optimal noise conditions of the LDs (particularly at high frequencies), we demonstrated that the NTL technology provides a viable solution for background reduction in CUPID.
}

\keywords{Double-beta decay, Cryogenic detectors, Scintillators, scintillation and light emission processes (solid, gas and liquid scintillators), Photon detectors for UV, visible and IR photons (solid-state), Pulse-shape discrimination methods}

\begin{document}
\maketitle
\flushbottom

%=========================================================
\section{Introduction}
\label{sec:intro} 

Low-temperature detectors (often referred to as cryogenic bolometers) represent a powerful method in particle detection to study neutrinoless double-beta ($0\nu\beta\beta$) decay, i.e., a hypothetical quasi-simultaneous conversion of two neutrons into two protons emitting only two electrons \cite{Agostini:2023,GomezCadenas:2023,Bossio:2023}. This process serves as a sensitive probe of total lepton number conservation and provides means to investigate fundamental neutrino properties, making it critically important for particle physics and cosmology. The most sensitive current bolometric search for $0\nu\beta\beta$ decay is CUORE (Cryogenic Underground Observatory of Rare Events) \cite{Adams:2022a,Alduino:2019}, located at the Gran Sasso underground laboratory in Italy. CUORE investigates the isotope $^{130}$Te using an array of $\sim$1000 TeO$_2$ bolometers, each measuring 50 $\times$ 50 $\times$ 50~mm, with a total mass of approximately 750 kg. CUORE sensitivity is currently limited by background events in the region of interest (ROI), primarily caused by $\alpha$ particles from radioactive surface contamination of the crystals and surrounding passive materials \cite{Adams:2024b}.

CUPID (CUORE Upgrade with Particle IDentification) \cite{Alfonso:2025cupid} is a proposed large-scale next-generation experiment, representing the natural progression of the CUORE project to significantly reduce the background level observed in CUORE. CUPID will deploy $\sim$1600 cryogenic calorimeters based on enriched Li$_{2}${}$^{100}$MoO$_4$ crystals 
and $\sim$1700 light detectors (LDs) in the existing infrastructure of CUORE. 
The LD, a separate cryogenic calorimeter, is used to capture scintillation signals from the Li$_2$MoO$_4$ crystal scintillators, enabling particle identification \cite{Pirro:2005ar,Poda:2021}. This particle identification capability suppresses CUORE's dominant surface $\alpha$-particle background in the ROI by at least two orders of magnitude, exploiting the much lower scintillation light produced by $\alpha$'s with respect to $\beta$'s and $\gamma$'s. The higher $Q_{\beta\beta}$ value of $^{100}$Mo (3034 keV vs. $^{130}$Te's 2528 keV) provides additional background mitigation by positioning the ROI above 2615 keV, the endpoint energy of prevalent $\gamma$-ray emissions from natural radioactive decay chains. 
A muon veto will be constructed around the CUORE cryostat to reduce the muon-induced background in CUPID. Thanks to these upgrades relative to CUORE, CUPID is projected to reach a background index in the ROI at the level of $\sim$10$^{-4}$ counts/keV/kg/yr (ckky) and to probe the existence of $0\nu\beta\beta$ decay with an order of magnitude higher sensitivity than current-generation experiments \cite{Alfonso:2025sensitivity}. 

The application of scintillating thermal detectors in view of CUPID has recently been demonstrated by two small-scale experiments\footnote{Another two small-scale demonstrators, AMoRE-pilot \cite{Alenkov:2019jis,Agrawal:2024} and AMoRE-I \cite{Agrawal:2025amoreI}, were realized for AMoRE next-generation experiment \cite{Alenkov:2015} with $^{100}$Mo-enriched scintillating low-temperature detectors instrumented with metallic magnetic calorimeters.}: CUPID-Mo~\cite{Armengaud:2020a,Augier:2022} and CUPID-0~\cite{Azzolini:2018tum,Azzolini:2019tta}. While both experiments followed the same underlying rationale, CUPID-0 used Zn$^{82}$Se crystals enriched in $^{82}$Se, with a $Q$-value of 2998 keV, while CUPID-Mo utilized Li$_2$MoO$_4$ crystals. These demonstrations validated the feasibility of the approach for future large-scale applications.

Following this promising approach, another two small-scale demonstrators --- CROSS (Cryogenic Rare-event Observatory with Surface Sensitivity)~\cite{Bandac:2020} and BINGO (Bi-Isotope $0\nu\beta\beta$ Next-Generation Observatory)~\cite{Armatol:2024bingo} --- along with two proposed next-generation experiments --- CUPID (CUORE Upgrade with Particle Identification)~\cite{Alfonso:2025cupid} and AMoRE (Advanced Mo-based Rare process Experiment)~\cite{Kim:2023} --- primarily focus on the isotope $^{100}$Mo embedded in Li$_2$MoO$_4$ bolometers for the search for $0\nu\beta\beta$ decay. CROSS, BINGO, and CUPID utilize the same type of temperature signal readout technology based on neutron transmutation doped (NTD) Ge sensors for the Li$_2$MoO$_4$ crystals and the LD Ge wafers. The work presented in this paper is primarily concerned with the challenges and developments related to these three experiments. In contrast to CUPID-Mo, which used cylindrical Li$_{2}${}$^{100}$MoO$_4$ crystals ($\oslash$44 $\times$ 45~mm), CROSS, BINGO, and CUPID will employ cubic-shaped crystals (45~mm side) to enhance packing within the experimental volume, increasing the detector mass and the efficiency of multi-crystal coincidences. 

CROSS is a project aimed at developing metal-coated bolometer technology capable of identifying near-surface interactions \cite{Bandac:2020,Bandac:2021}, as a potential solution for further reduction of the background originating from surface contamination. Following the purification and crystallization protocols developed by LUMINEU~\cite{Berge:2014,Armengaud:2017,Grigorieva:2017} and applied for CUPID-Mo \cite{Armengaud:2020a}, 32 cubic Li$_{2}${}$^{100}$MoO$_4$ crystals (97\% enrichment) have been produced for the CROSS experiment and used in several tests at the Canfranc underground laboratory (LSC) in Spain \cite{Armatol:2021b,CrossCupidTower:2023a,CROSSdetectorStructure:2024}. 

BINGO (Bi-Isotope $0\nu\beta\beta$ Next-Generation Observatory)~
is a project focused on developing and testing innovative technologies for $0\nu\beta\beta$ decay experiments based on scintillating bolometers, with particular emphasis on background reduction. Key advancements include the use of an innovative holder design, where LDs perform as active shields for most passive materials, and the implementation of an internal cryogenic veto. Among the technologies adopted by BINGO, the enhancement of LDs through the Neganov-Trofimov-Luke (NTL) effect \cite{Neganov:1985,Luke:1988} --- central to this article --- is particularly noteworthy. This approach has also been extended to CROSS, which initially did not include LDs \cite{Bandac:2020}, and CUPID, which initially envisioned ordinary LDs as in CUPID-0 and CUPID-Mo \cite{CUPIDInterestGroup:2019inu}.

A mechanical structure adopted by CUPID \cite{Alfonso:2022}, so-called the CUPID Baseline structure, introduces an innovative design that simplifies the assembly of a large-scale detector while maintaining stringent low-background requirements essential for rare-event searches. In this configuration, scintillating crystals and copper frames holding the LDs are stacked without individual direct thermal connections between the crystals and the copper heat sink. This arrangement mitigates potential issues arising from differential thermal contractions between the various materials used. The novel design was validated through tests with a CUPID-like tower,  
which comprises 28 Li$_2$MoO$_4$ scintillating bolometers, showing good detector thermalization and performance \cite{Alfonso:2025gdpt}. However, the average noise of LDs was found to be a factor 2  higher than the CUPID requirement (noise RMS $<$ 150 eV) and early results of prototype tests \cite{Alfonso:2022,Alfonso:2023}. To address this issue, the detector design was slightly modified 
and a test of the second CUPID prototype tower is planned in 2025 \cite{Alfonso:2025cupid}.

The detector structures used for CROSS \cite{CROSSdetectorStructure:2024} and BINGO \cite{Armatol:2024bingo} are based on different designs that focus mainly on the direct thermalization of each bolometer to its copper support which in usually beneficial for detector performances. Both designs were developed also as an alternative design for the structure of the CUPID holder in the framework of a risk mitigation strategy. 
Notably, the refined holders achieve a lower Cu-to-Li$_2$MoO$_4$ mass (CROSS) or surface area (BINGO) ratio than any other macrobolometer holder implemented to date, reducing background contributions from surface and nearby materials. Extensive tests conducted above ground at IJCLab (Orsay, France) and underground at LSC (Laboratorio Subterr\'aneo de Canfranc, Spain) \cite{CROSSdetectorStructure:2024,Armatol:2024bingo} confirmed that both Li$_2$MoO$_4$ bolometers and Ge LDs meet performance standards required for CROSS, BINGO and potentially CUPID experiments.

As mentioned earlier, BINGO, CROSS, and CUPID aim to enhance the performance of LDs by leveraging the NTL effect, which amplifies thermal signals in the presence of an electric field~\cite{Neganov:1985,Luke:1988} generated by a set of electrodes deposited on the LD Ge wafer~\cite{Novati:2019}. The potential energy of electron-hole pairs produced by light absorption is converted into heat during their drift toward the electrodes, resulting in a significant improvement in the signal-to-noise ratio (SNR) of the light signal --- up to a factor 10.
We are optimizing the electrode configuration for square 45$\times$45~mm LDs, such as those used in BINGO and CROSS. This configuration can be readily adapted to the CUPID Baseline structure, which utilizes octagonal LDs with a shape and size closely resembling those of BINGO and CROSS.

NTL-operated LDs are highly effective for controlling random-coincidence background induced by $^{100}$Mo two-neutrino double-beta ($2\nu\beta\beta$) decay events ---one of the major background components in $0\nu\beta\beta$ ROI of a large-scale experiment with $^{100}$Mo-enriched thermal detectors (like CUPID \cite{Alfonso:2025cupid}) due to their slow response and a comparatively high $2\nu\beta\beta$ rate \cite{Chernyak:2012,Chernyak:2014}. The comparatively fast rise time and large signal amplitude of NTL-assisted LDs improve pile-up rejection through pulse-shape discrimination (PSD)~\cite{Chernyak:2017,CROSSpileup:2023}. 

Given the importance of NTL technology for LDs in BINGO, CROSS and CUPID, we conducted a test of 10 identical prototypes of these devices. These were coupled to 45-mm-side cubic Li$_2$MoO$_4$ and TeO$_2$ crystals using the CROSS detector structure (Sect.~\ref{sec:Prototype_Tower}). The array was operated within the CROSS setup using a pulse-tube cryostat at the LSC (Sect.~\ref{sec:Cryo_test}), where we extensively evaluated detector performance (Sect.~\ref{sec:Results}). Based on these results, we investigated pile-up rejection capabilities across several detector configurations relevant to CROSS, BINGO, and CUPID (Sect.~\ref{sec:Prospects}). This article provides a detailed account of detector fabrication, tower assembly, low-temperature testing, performance results, and prospects for application in CUPID.

%==============================================================================

\section{Ten-crystal prototype tower of CROSS}
\label{sec:Prototype_Tower}

%----------------------------------------------------------------
\subsection{Mechanical structure of a detector array} 

The new structure, recently developed and tested by the CROSS collaboration \cite{CROSSdetectorStructure:2024} in preparation for $^{100}$Mo $0\nu\beta\beta$ decay searches \cite{Bandac:2020}, has several features that are essential for large-scale $2\beta$-search bolometric experiments:

\begin{itemize} 

\item The structure is designed to hold crystals of $45\times45\times45$~ mm, that is, the same size as the detector chosen by CUPID \cite{Alfonso:2025cupid}. Furthermore, the crystals are accompanied by CUPID-like Ge wafers ($45\times45\times0.3$~mm) used as bolometric photodetectors. 
\item No specific holder is foreseen for Ge wafers, which are mounted in the common structure with the crystals. This key feature allows the reduction of the amount of passive materials in the design. 
\item Highly radiopure elements, Cu frames and 3D-printed PLA (polylactic acid) spacers, with a very low mass budget relative to the detector module weight were used in the construction. Reducing the mass and surface area ratios of passive detector materials is essential for reducing the surface-derived background component, which is one of the main contributions to the background budget in the ROI of bolometric experiments. The structure design includes two options for the Cu content, namely \emph{Thick} and \emph{Slim} versions with a Cu/Li$_2$MoO$_4$ mass ratio of around 15\% and 6\%, respectively, and the same amount of spacers (PLA/Li$_2$MoO$_4$ mass ratio is around 0.4\%). 
\item Each absorber material has direct heat sinking through supporting elements made of 3D-printed PLA spacers and clips. The spacers between Li$_2$MoO$_4$ and Ge were adapted to avoid heat crosstalk between the detectors. Consequently, the thermal network of each detector module is independent of the others, as is typically used in past and present bolometric detectors for $2\beta$ decay searches. 
\item A single detector unit of the structure consists of two crystals and two Ge wafers, providing the capability of modular tower construction, similar to what is proposed for CUPID \cite{Alfonso:2022}. (Note: 13 two-crystal modules per tower can fit the height of the CUORE cryostat's experimental volume, compared to 14 ones of the CUPID Baseline structure.) 
\item The detectors are instrumented with the same type of temperature sensors, NTD Ge thermistors \cite{Haller:1994}, foreseen in CUPID. The thermal coupling to the absorber materials is done with a new kind of glue (UV-cured), providing similar or better performance than the ``classical'' widely used two-component epoxy glue initially considered for the thermistors' gluing in CUPID. Note that the structure allows for wire bonding of the thermistors (accessible laterally) after the whole assembly of a tower. 
\item Detector performance ---particularly energy resolution and particle identification capability--- compatible with CUPID demands has been demonstrated in underground measurements in a pulse-tube cryostat \cite{CROSSdetectorStructure:2024}, adopting the same technology as the CUPID facility and thus subject to a similar vibrational noise environment \cite{Olivieri:2017}. 

\end{itemize} 

For the present work, we adopted the \emph{Slim} version of the CROSS mechanical structure to demonstrate both a risk-mitigating alternative to the CUPID Baseline detector structure and the technology of bolometric LDs with NTL signal amplification, selected for the final CUPID detector design \cite{Alfonso:2025cupid}.

%----------------------------------------------------------------
\subsection{10-crystal demonstrator} 

For the construction of the demonstrator, we used ten crystals, each with dimensions of $45\times45\times45$~mm. Six of these crystals are lithium molybdates, with two of them enriched in $^{100}$Mo to a level of 97.5\%. The remaining four crystals are tellurium dioxide, enriched in $^{130}$Te to a level of 91.4\%. All Li$_2$MoO$_4$ samples with a natural isotopic composition of Mo were produced by the RMD company (Watertown, MA, USA) using the Czochralski crystal growth method as part of R\&D activities for CUPID \cite{RMD_LMO_bolometers:2025}. Two Li$_2$MoO$_4$ crystals containing molybdenum enriched in $^{100}$Mo belong to a batch of 32 samples produced for the CROSS experiment using the low-temperature-gradient Czochralski technique and were used as reference detectors in the mechanical structure tests of CROSS \cite{CROSSdetectorStructure:2024}. The last four massive crystals of the tower were randomly selected from a batch of six $^{130}$TeO$_2$ crystals recently developed for the CROSS experiment using highly purified tellurium powder enriched in the $^{130}$Te isotope and the Czochralski crystal growth method \cite{CROSS_enriched_TeO:2024}. 

\begin{figure}
\centering
\includegraphics[width=0.9\textwidth]{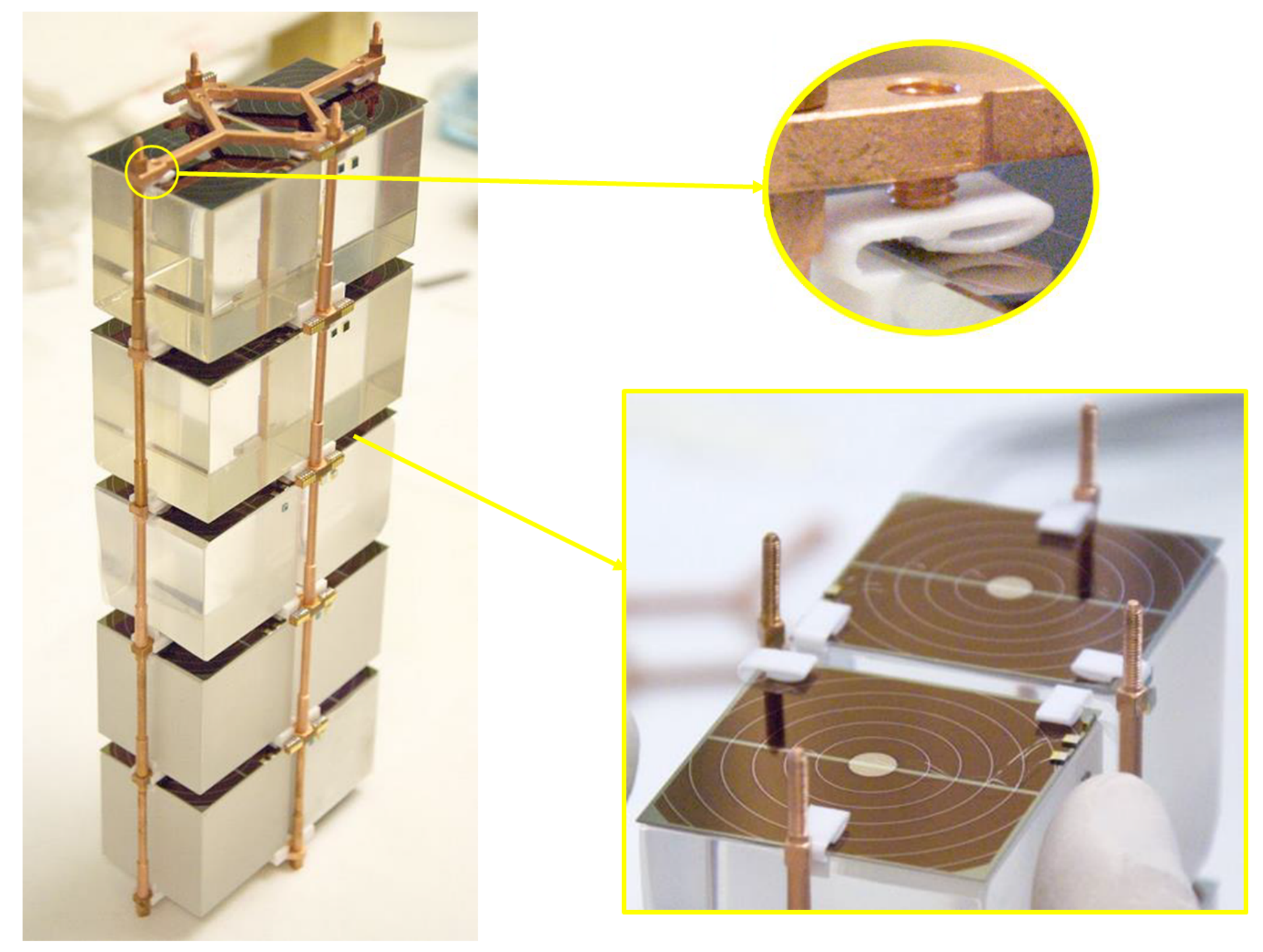}
\caption{(Left) A 10-crystal tower containing 45-mm side cubic crystals; natural Li$_2$MoO$_4$ samples are used on the first two floors from the top, $^{100}$Mo-enriched Li$_2$MoO$_4$ crystals are in the middle, and the last two floors are occupied by $^{130}$Te-enriched TeO$_2$ samples. 
Two key upgrades of the CROSS detector structure \cite{CROSSdetectorStructure:2024} to improve light detector performance are the enlarged 3D-printed PLA clips (Right top) to improve LD's clamping and a set of Al electrodes deposited on Ge wafers (Right bottom) to exploit thermal signal amplification based on the Neganov-Trofimov-Luke effect in semiconductors. Two small (2 $\times$ 2 $\times$ 0.5 mm) Al-coated Si transition pieces were glued on each Ge wafer to provide electrical connection from contacts on the Kapton to Al electrodes.}
\label{fig:Tower_ijclab}
\end{figure}

Before assembly, we attached an NTD-Ge thermistor (with a size of 3 $\times$ 3 $\times$ 1 mm) and a Si:P heater \cite{Andreotti:2012} to each crystal. The coupling materials used were UV-curable glue (PERMABOND\textregistered ~620) and bi-component epoxy (Araldite\textregistered ~Rapid), respectively. Using components of the \emph{Slim} mechanical structure, we mounted two crystals per floor, with a total of five floors. At each floor, both crystals were placed on the Y-shaped part of the Cu frame; thermal decoupling was secured by 3D-printed PLA spacers designed also to confine the crystal position laterally. Then, we placed a square-shaped Ge wafer (described in detail in the next section), equipped with a smaller NTD sensor (approximately 1 $\times$ 3 $\times$ 1 mm), on top of each crystal. Three 3D-printed PLA clamps were developed to avoid full-contact coupling and apply a force towards the crystal with a tiny copper screw. Based on prototype tests, we expect improved performance of LDs compared to the predecessors studied in the early stages with previous versions of the clamps \cite{CROSSdetectorStructure:2024}. Finally, an identical Cu frame is placed above, and the construction is secured using four Cu columns and nuts. The procedure is then repeated to build a 5-floor tower containing ten large-volume ($\sim$90 cm$^3$) bolometers and 10 Ge LDs; the full tower and some parts of a single detector module are shown in figure \ref{fig:Tower_ijclab}. 

%----------------------------------------------------------------
\subsection{Light detectors with thermal signal amplification} 

In contrast to the ``standard'' bolometric Ge LDs used in the CROSS mechanical structure validation tests \cite{CROSSdetectorStructure:2024}, we decided to modify these devices to enhance their SNR by using the thermal signal amplification induced by the NTL effect in semiconductors, which is briefly described in this section.

%----------------------------------------------------------------
\subsubsection{NTL-effect-based devices in a nutshell} 

A bolometer exploiting the NTL effect can be conceptualized as follows. A semiconductor absorber is equipped with electrodes on its surface (or even contactless) to establish an electric field within the absorber volume. Similar to classical semiconductor detectors, this field enables the collection of charges ---consisting of electron-hole pairs--- generated by particle interactions (e.g., absorbed scintillation light). As these charge carriers drift towards the electrodes under the influence of the electric field, they deposit additional heat in the crystal through the Joule effect, which is then measured by a phonon sensor. With a sufficiently high applied voltage, the heat produced by charge motion exceeds that initially generated by particle interaction, becoming the predominant component of the thermal signal. Essentially, such a bolometric device operates as an ionization detector with a heat-based charge readout (i.e., a charge-to-heat transducer). The amplification of the thermal signal is given by the following formula:

\begin{equation}
    E_{NTL} = E_0 \cdot (1 + e \cdot V_{NTL} \cdot \eta / \epsilon ) = E_0 \cdot G_{NTL},
\label{eq:NTL_gain}
\end{equation}

\noindent where $E_0$ is the initial thermal energy deposited by particle interaction, $E_{NTL}$ is the thermal energy measured by the phonon sensor after the NTL amplification, $V_{NTL}$ is the voltage across two nearby electrodes, $\epsilon$ is the energy necessary to create an electron-hole pair in the absorber 
(for germanium at cryogenic temperatures, it is equal to around 3 eV \cite{Antman:1966} for the incident radiation with energy above a few eV, while it becomes energy-dependent for excitations with lower energies \cite{Koc:1957}) 
and $\eta$ is a coefficient ranging from 0 to 1 that provides the fraction of electron-hole pairs which actually contribute to the NTL amplification. 
In practice $\eta$ is expected to be dependent on surface and bulk material properties related to charge trapping and impact ionization and can vary for different particle interaction topologies and semiconductor purities as modeled in \cite{Wilson:2024}. 
The case of $\eta$ = 1 corresponds to an idealized scenario without trapping or impaction ionization. 
The second equality in Eq. \ref{eq:NTL_gain} defines the NTL amplification $G_{NTL}$, which can be approximated by $e \cdot V_{NTL} \cdot \eta / \epsilon$ since this term is much greater than 1 in all practical cases. Notice that the term $E_0 / \epsilon$ provides the number of electron-hole pairs initially generated. If they are fully collected, they contribute to the total heat released with their initial potential energy, that is $e \cdot V_{NTL}$ for a single pair.

Eq. \ref{eq:NTL_gain} suggests that an arbitrarily large amplification can be obtained by increasing $V_{NTL}$ indefinitely. In reality, there are two limitations to be taken into account:
(1) after a certain threshold, $V_{NTL}$ cannot be increased without developing a destructive leakage current that heats the entire device and sometimes the whole cryostat; 
(2) $\eta$ is always less than 1, 0.3--0.5 in the present Ge light detector design \cite{Novati:2019}. A subtler limitation concerns noise: low-frequency noise sometimes appears before the maximum applicable NTL voltage is reached, limiting the SNR despite increasing pulse amplitudes \cite{Novati:2019}. However, amplification can be substantial enough that a single electron-hole pair can produce a detectable heat pulse above baseline noise, making the LD nearly single-photon sensitive \cite{Poda:2021}. 

In addition to the above, we note that one of the main advantages of the NTL technology for CUPID is that the temperature readout scheme of the LD is unchanged concerning the initial CUPID configuration \cite{CUPIDInterestGroup:2019inu}. The same NTD Ge thermistors can be used, implying that no change is required in the room-temperature Si-JFET-based front-end electronics, which have been successfully and extensively used in CUORE to read out NTDs. The only modifications, a part of the increased LD complexity, are related to additional cabling and stable low-noise power supplies to apply the NTL voltage, to the installation of optic fibers for a LED-driven regeneration (the neutralization of space charges) of such devices, and to extra care about IR shielding of the experimental volume.

%----------------------------------------------------------------
\subsubsection{Fabrication of NTL-assisted light detectors} 

For the demonstration test, we fabricated 10 identical Ge LDs based on 45 $\times$ 45 mm square slices (as used in CROSS and BINGO, and close to the final CUPID size), obtained by cutting 4''-diameter 0.3-mm-thick Ge wafers (figure \ref{fig:Tower_ijclab}). The purity of these wafers in terms of the maximum net electrical active impurity concentration is in the range (1--3) $\times$ 10$^{10}$ cm$^{-3}$. We notice that the shadow masks we used in the process of Al electrode evaporation on the 10 Ge wafers are conceived for circular samples, such as the ones developed in \cite{Novati:2019} and are not adapted to square slices (figure \ref{fig:Tower_ijclab}, right bottom). Therefore, only 56\% of the surface of the square LD participates in the NTL amplification when the voltage is applied to the electrodes.

The electrode deposition scheme follows an established recipe \cite{Novati:2019}, developed at IJCLab (Orsay, France) and applied there for thin-film deposition. The electrode fabrication is performed with a dedicated electron-beam evaporator in a clean room, following a procedure that must be meticulously implemented to avoid dangerous leakage currents resulting to the partial warming-up of a dilution refrigerator. The main steps are summarized below.

\begin{itemize}

    \item Definition of electrode-patterned structures in concentric rings through shadow masks, with a typical electrode pitch of 4 mm.
    
    \item Ar-ion bombardment – Ar is ionized by electron beam and ions are accelerated onto the wafer under 90 V.
    
    \item Formation of an amorphous 50-nm-thick layer of Ge and H (to saturate Ge dangling bonds) – similar procedure as Ar bombardment – H-ions are accelerated under 80 V.
    
    \item Deposition by evaporation of 100-nm-thick 0.2-mm-wide Al electrodes by electron gun, according to the pattern defined by the shadow masks.
    
    \item Increase of electrode thickness up to 160 nm in some parts for ultrasonic bonding, necessary for electrical contact with the rings.
    
    \item Coating of the whole wafer (but the bonding zone) by a 70-nm-thick SiO antireflective layer by thermal evaporation, heating the source to 1000$^\circ$C.
    
%    \item 
\end{itemize}

Therefore, NTL electrodes in their simplest form consist of a pattern of concentric Al rings, which are electrically connected by ultrasonic Al-wire bonding with an alternate scheme, so that the same NTL voltage is applied across each couple of adjacent rings (figure \ref{fig:Tower_ijclab}, right bottom).

%==============================================================================
\section{Test of the 10-crystal array at the Canfranc underground laboratory} 
\label{sec:Cryo_test}

%----------------------------------------------------------------
\subsection{Detector installation in the CROSS cryogenic facility} 

\begin{figure}
\centering
\includegraphics[width=0.45\textwidth]{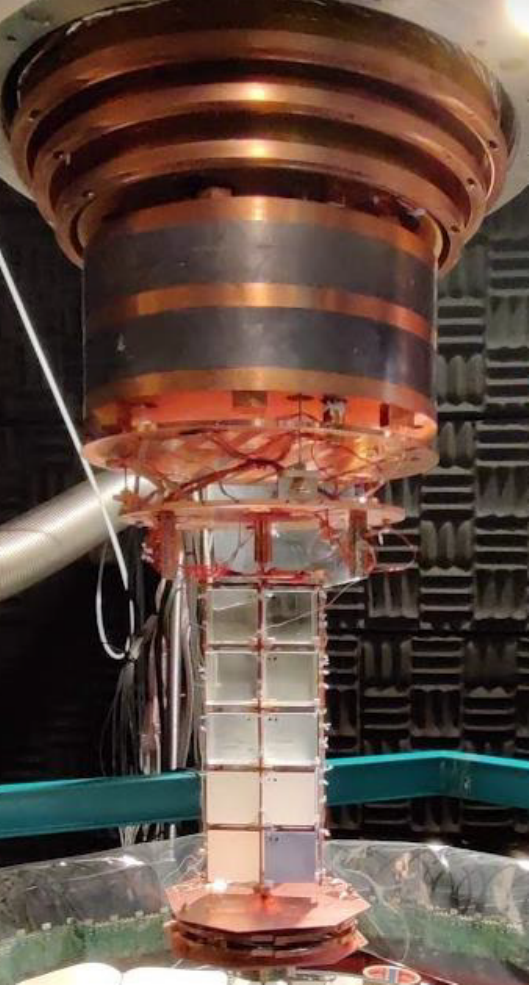}
\includegraphics[width=0.4\textwidth]{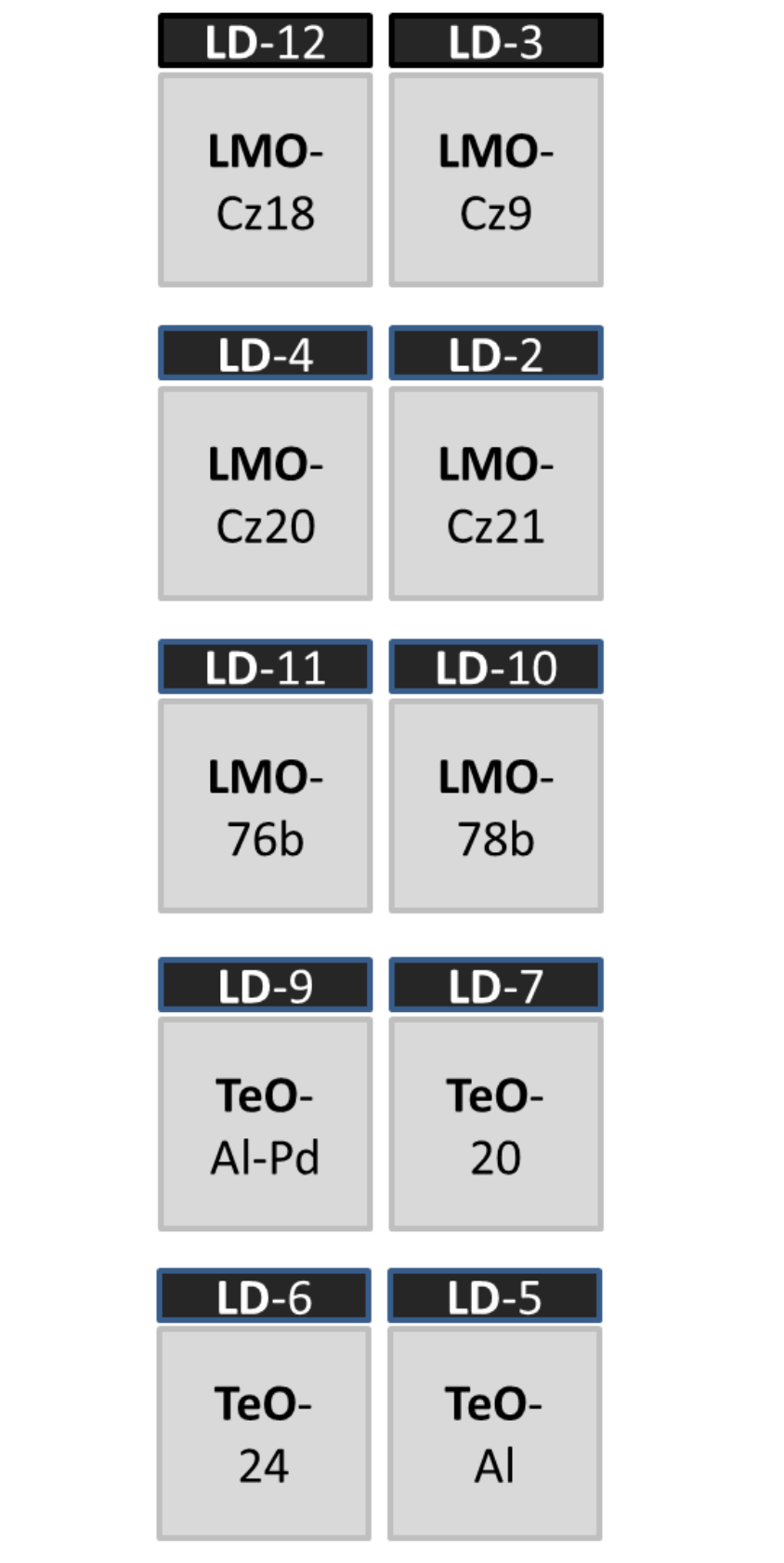}
\caption{(Left) View of the 10-crystal tower coupled to the copper detector plate within a pulse-tube dilution refrigerator. The image showcases the main components of the internal shielding, consisting of copper-sandwiched lead, and the joints of the copper thermal screens positioned above the tower. (Note: An additional R\&D detector array \cite{SURFACE_Si:2025}, not discussed in this study, was connected to the lower section of the 10-crystal tower.) (Right) Schematic representation of the channel mapping for the 10-crystal array of bolometric detectors, illustrating the spatial arrangement and connectivity of the individual sensing elements. A single module (LD-6 \& TeO-24) was not operational due to a readout-related issue.}
\label{fig:Tower_lsc}
\end{figure}

The 10-crystal tower assembled at IJCLab (France) was then transported to the LSC in Spain and mounted on the detector plate inside a CROSS-dedicated dilution refrigerator, as shown in figure \ref{fig:Tower_lsc} (left).

The CROSS facility \cite{Olivieri:2020,CROSS_Magnetic_dampers:2023} consists of a pulse-tube-based cryostat by Cryoconcept (France) with a large experimental volume ($\oslash$60 $\times$ 80 cm) capable of reaching $\sim$10 mK. The cryostat is equipped with the Ultra-Quiet Technology{\texttrademark} (UQT) \cite{UQT} to mechanically decouple the pulse tube from the dilution unit of the cryostat, mitigating pulse-tube-induced vibrational noise \cite{Olivieri:2017}. 
Moreover, to further reduce the transmission of vibrational noise, a system of 9 springs with 3 magnetic dampers at the 1K stage has been installed to suspend the detector plate \cite{CROSS_Magnetic_dampers:2023}.
The experimental volume is shielded against all the parts of the cryostat, including the mixing chamber, by two layers of Cu and low-radioactivity Pb. The lateral and bottom parts of the experimental volume are shielded by 25-cm-thick lead (8.5 t of the total weight). The lead can be classified as a low-radioactivity type regarding uranium and thorium content, however it presents a notable activity of $^{210}$Pb ($\sim$60 Bq/kg), not affecting the background for $0\nu\beta\beta$ searches. A stainless steel vessel is constructed around the external Pb shield to flush the area around the outer vacuum chamber of the cryostat by deradonized air flow. 

The facility's data acquisition system is based on room-temperature low-noise voltage-sensitive DC electronics implemented on 12-channel cards with 24-bit ADC to acquire the thermistors' voltage output as a continuous data stream \cite{Carniti:2020,Carniti:2023}.

%----------------------------------------------------------------
\subsection{Data taking and processing} 

The channels' mapping is shown in figure \ref{fig:Tower_lsc} (right). Due to a readout problem, unrelated to the intrinsic properties of the devices, a single LD (channel LD-6) and a massive bolometer (TeO-24) were unusable at low temperatures. 
We performed two measurement campaigns with two different base temperature: 17 mK (referred to as ``colder'')  and 22 mK (referred to as ``warmer''). These two configurations allowed for the LD optimization first prioritizing SNR (``colder'' configuration) and secondly time response (``warmer'' configuration). 
A $^{232}$Th source made of a thoriated tungsten wire was inserted inside the Pb shield to calibrate the detectors. A sampling rate of 2 kHz / 10 kHz and a cut-off frequency of the active Bessel filter of 0.3 kHz / 2.5 kHz were set, for the data taking at 17 mK / 22 mK. 

The optimal working points (corresponding to NTD bias currents that maximize SNR) in the ``colder'' measurements were defined channel-by-channel by scanning the amplitude of heater-/LED-induced pulses and the baseline noise with the change in the NTD current. To further reduce the rise times of the LD thermal signals at the ``warmer'' operational temperature, we polarized the NTDs of these devices at a bias current of 10 nA, an order of magnitude higher than the usual ones that aim at maximization of SNR. 

The acquired data were processed with the help of a MATLAB-based tool \cite{Mancuso:2016} to filter the data (reducing noise) and to compute signal amplitudes (proportional to energy) and several PSD parameters for each triggered event. 
In particular, the PSD parameters relevant for this study are rise time ($\tau_{rise}$) and decay time ($\tau_{decay}$) computed in the 10\%--90\% of the leading part and 90\%--30\%  of the descending part of a signal, respectively. 
The data processing program exploits the Gatti-Manfredi optimum filter \cite{Gatti:1986}, which provides the best estimation of the signal amplitude based on the signal shape and the noise power spectrum. To build the filter's transfer function, we construct a data-driven template of a signal (an average of tens-hundreds of high-energy events) and of the noise power density (an average of 20000 waveforms with no signal) for each channel and working point.

%==============================================================================
\section{Results and discussion} 
\label{sec:Results}

%----------------------------------------------------------------
\subsection{Performance of reference Li$_2$$^{100}$MoO$_4$ bolometers} 

Two Li$_2$$^{100}$MoO$_4$ bolometers used in the 10-crystal tower were earlier tested in the CROSS setup \cite{CROSSdetectorStructure:2024}), thus, both are considered in the present work as representative (reference) massive thermal detectors of the array. Low-temperature characterizations of other crystals --- natural Li$_2$MoO$_4$ and enriched $^{130}$TeO$_2$ --- are the topics of dedicated studies (\cite{RMD_LMO_bolometers:2025} and \cite{CROSS_enriched_TeO:2024}, respectively), and are therefore omitted in the present work (the only exception is the Li$_2$MoO$_4$ scintillation light yield used for the LD calibration in the NTL mode as briefly described in section \ref{sec:LY_calibration}). 

\begin{figure}
\centering
\includegraphics[width=0.9\textwidth]{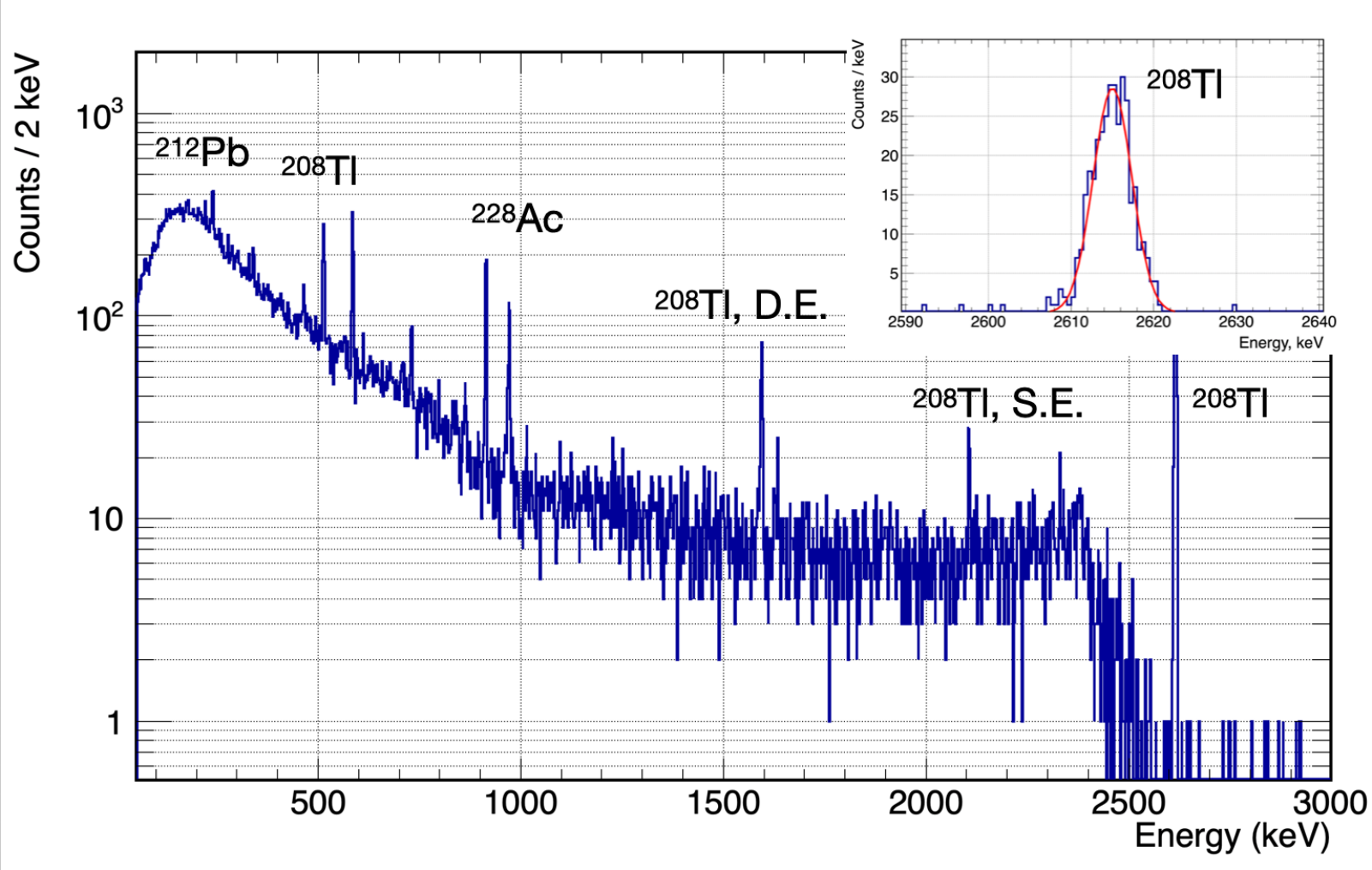}
\caption{Energy spectrum of a $^{232}$Th source measured with the reference Li$_2$$^{100}$MoO$_4$ bolometer (Li$_2$MoO$_4$-76b) in the CROSS underground setup (17 mK data, 83 h). The most intense $\gamma$-ray peaks observed in the spectrum are labeled (D.E. and S.E. are double and single escape peaks, respectively). A Gaussian fit to the 2615 keV $\gamma$ peak of $^{208}$Tl is shown in the inset; the energy resolution is 5.7(3) keV FWHM.}
\label{fig:Th_spectrum}
\end{figure}

The performances of both reference Li$_2$$^{100}$MoO$_4$ bolometers, reported in table \ref{tab:Li$_2$MoO$_4$_ref_performance}, are compatible with the expected value for the CROSS facility \cite{CROSSdetectorStructure:2024}.
Being polarized at a few nA currents at 17 mK, both bolometers show a similar high sensitivity, i.e., a comparatively high voltage signal (about 80 nV) is developed by a 1 keV thermal signal. The higher NTD resistance of Li$_2$MoO$_4$-76b is compatible to what has been observed for a higher stress on the glued sensor \cite{CROSSdetectorStructure:2024}. 
The energy resolution of the baseline noise, 1--2 keV FWHM, is among the best reported for Li$_2$MoO$_4$-based bolometers tested in the CROSS setup \cite{Armatol:2021b,CrossCupidTower:2023a,CROSS_Magnetic_dampers:2023,CROSSdeplLMO:2023,CROSSdetectorStructure:2024} and worldwide \cite{Bekker:2016,Armengaud:2017,Armengaud:2020a,Armatol:2021a,Alfonso:2022,Alfonso:2025gdpt}. 
As expected, both detectors show a good energy resolution over the wide range of energy spanned by 
$\gamma$ quanta emitted by a $^{232}$Th calibration source, as illustrated in figure \ref{fig:Th_spectrum}. In particular, the energy resolution of the 2615~keV $\gamma$ quanta of $^{208}$Tl, near the ROI for $^{100}$Mo $0\nu\beta\beta$ decay (3034 keV), is around 6 keV FWHM. 
It is important to note that the reference (and all other) crystals of the present work were mounted in the finer (\emph{Slim}) version of the CROSS structure in contrast to previous tests of these two samples in the thicker (\emph{Thick}) design \cite{CROSSdetectorStructure:2024}. This result proves that both designs of the CROSS detector assembly allow to obtain high performance. 

\begin{table}
\centering
\caption{Best performance of the CROSS reference Li$_2$$^{100}$MoO$_4$ bolometers tested at 17 mK. We report the parameters of the chosen working points: NTD resistance $R_{NTD}$ at a given current $I_{NTD}$, detector's sensitivity ($A_{signal}$), energy resolution (full width at the half maximum) measured for zero energy deposition (FWHM$_{baseline}$) and for 2615 keV $\gamma$ of $^{208}$Tl (FWHM$_{2615}$). Statistical uncertainties of $R_{NTD}$ and $A_{signal}$ values are below the given precision.}
\smallskip
\begin{tabular}{cccccc}
\hline
Detector  & $R_{NTD}$ & $I_{NTD}$ & $A_{signal}$ & FWHM$_{baseline}$ & FWHM$_{2615}$  \\
ID & (M$\Omega$) & (nA)   & (nV/keV) & (keV) & (keV)  \\
\hline
\hline
Li$_2$MoO$_4$-76b    & 4.8  & 2.0  & 84 & 1.10(2) & 5.7(3)  \\ % 
\hline
Li$_2$MoO$_4$-78b    & 1.8 & 3.0 & 81 & 1.71(5) & 5.9(10)  \\ %
\hline
\end{tabular}
\label{tab:Li$_2$MoO$_4$_ref_performance}
\end{table}

%----------------------------------------------------------------
\subsection{Characterization of bolometric Ge light detectors} 

\begin{table}
\centering
\caption{Best performance of bolometric Ge light detectors operated in the 10-crystal tower installed in the CROSS cryostat at the Canfranc underground laboratory.  For each channel, we report the working point of the NTD thermistor (resistance $R_{NTD}$ at a given current $I_{NTD}$), pulse shape time constants (rise- and decay-time parameters, $\tau_{rise}$ and $\tau_{decay}$), detector sensitivity represented by a signal voltage amplitude per unit of the deposited energy ($A_{signal}$), and RMS baseline width ($\sigma_{baseline}$) after application of the optimum filter.  Statistical uncertainties of $R_{NTD}$ values are below the given precision. The widths of $\tau_{rise}$ and $\tau_{decay}$ distributions are taken as uncertainties of the quoted parameters. Uncertainties of $A_{signal}$ and $\sigma_{baseline}$ values are statistical only.
}
\smallskip
\begin{tabular}{lccccccc}
\hline
Detector  & $T_{plate}$ & $R_{NTD}$ & $I_{NTD}$ & $\tau_{rise}$ & $\tau_{decay}$ & $A_{signal}$ & $\sigma_{baseline}$ \\
ID & (mK) & (M$\Omega$) & (nA)   & (ms) &  (ms) & ($\mu$V/keV) & (eV)  \\
\hline
\hline
LD-2    & 17 & 4.6  & 2.0   & 1.54(24) & 6.1(3) & 1.49(1)  & 58(1)   \\ % m703
~       & 22 & 0.80 & 10    & 0.86(7)  & 4.4(1) & 0.65(2) & 78(1)   \\ % m808
\hline
LD-3    & 17 & 7.0  & 1.5   & 1.68(5)  & 3.1(3) & 1.97(3)  & 63(1)   \\ % m703
~       & 22 & 0.94 & 10    & 0.69(11) & 3.0(1) & 0.75(3) & 115(1)   \\ % m808
\hline
LD-4    & 17 & 12   & 1.0   & 2.12(33) & 3.7(3) & 2.07(1)  & 63(1)   \\ % m703
~       & 22 & 0.97 & 10    & 0.59(6)  & 3.3(1) & 0.56(1) & 97(3)   \\ % m808
\hline
LD-5    & 17 & 4.0  & 1.6   & 1.87(8)  & 9.6(3) & 1.83(3)  & 56(1)   \\ % m703
~       & 22 & 0.57 & 10    & 0.85(2)  & 5.7(1) & 0.64(1) & 72(1)   \\ % m808
\hline
LD-7    & 17 & 4.0  & 2.0   & 1.37(14) & 8.9(3) & 1.39(1)  & 146(2)   \\ % m703
~       & 22 & 0.81 & 10    & 0.59(2)  & 6.1(1) & 0.57(1) & 115(2)   \\ % m808
\hline
LD-9    & 17 & 8.5  & 2.0   & 1.49(9)  & 7.2(3) & 1.57(1)  & 63(1)   \\ % m703
~       & 22 & 0.88 & 10    & 0.64(2)  & 5.3(1) & 0.68(1) & 77(1)   \\ % m808
\hline
LD-10   & 17 & 8.4 & 1.5    & 1.66(3)  & 2.5(4) & 1.72(1)  & 79(1)   \\ % m703
~       & 22 & 1.1 & 10     & 0.53(3)  & 3.0(1) & 0.66(1) & 93(1)   \\ % m808
\hline
LD-11   & 17 & 12  & 1.0    & 1.75(13) & 3.3(2) & 1.93(3)  & 117(2)   \\ % m703
~       & 22 & 1.1 & 10     & 0.53(3)  & 3.4(2) & 0.66(1) & 108(1)   \\ % m808
\hline
LD-12   & 17 & 6.0 & 2.0    & 2.46(5)  & 2.3(2) & 1.02(2)  & 132(2)   \\ % m703
~       & 22 & 1.1 & 10     & 0.53(5)  & 1.9(1) & 0.47(1) & 159(2)   \\ % m808
\hline
\hline
Mean    & 17 & 6.3 & 1.5    & 1.72(8) & 4.0(3) & 1.59(1)  & 76(1)   \\ % m703
~       & 22 & 0.88 & 10    & 0.62(3) & 3.5(1) & 0.62(1) & 96(1)   \\ % m808
\hline
\end{tabular}
\label{tab:LD_performance}
\end{table}

\begin{figure}
\centering
\includegraphics[width=0.85\textwidth]{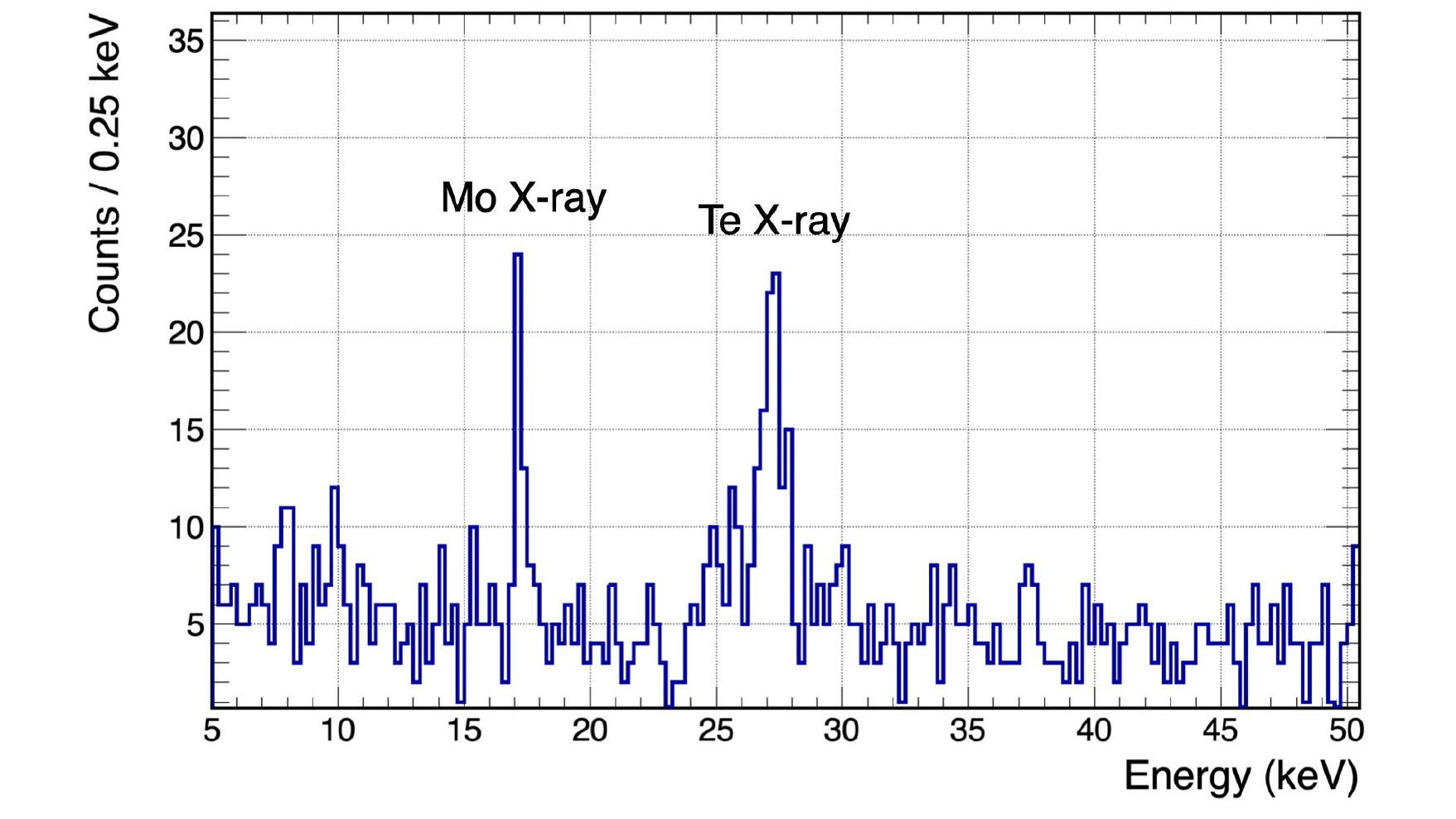}
\caption{Energy spectrum accumulated by a bolometric Ge light detector (LD-9) after 83 h of calibrations with a $^{232}$Th source at 17 mK. Two peaks present in the spectrum correspond to Mo and Te X-rays induced in the neighbor Li$_2$$^{100}$MoO$_4$ (Li$_2$MoO$_4$-76b) and $^{130}$TeO$_2$ (TeO-Al-Pd) crystals by $\gamma$ rays.}
\label{fig:LD_calibration}
\end{figure}

Bolometric photodetectors are key devices in the present work, and thus, we overview their performance, summarized in table \ref{tab:LD_performance}, in more detail. To assess the performance of LDs, we calibrated these devices with the X-ray fluorescence induced in the coupled crystals by the source-emitted $\gamma$ radioactivity. An example of an LD calibration spectrum is presented in figure \ref{fig:LD_calibration}.

At ``colder'' temperature conditions, we polarized the NTDs of LDs at currents of 1--2 nA (4--12 M$\Omega$ resistances), leading to a sensitivity of 1--2 $\mu$V/keV (with a mean of 1.6 $\mu$V/keV), typical for such devices in this setup \cite{CrossCupidTower:2023a,CROSSdetectorStructure:2024}. 
We observed baseline resolution values in the range 0.05--0.15 keV RMS (0.07 keV RMS mean), with a factor 2 improvement with respect to previous values measured with the \emph{Slim} design \cite{CROSSdetectorStructure:2024}. This improvement can be traced back to the use of enlarged 3D-printed spacers.  

\begin{figure}
\centering
\includegraphics[width=0.7\textwidth]{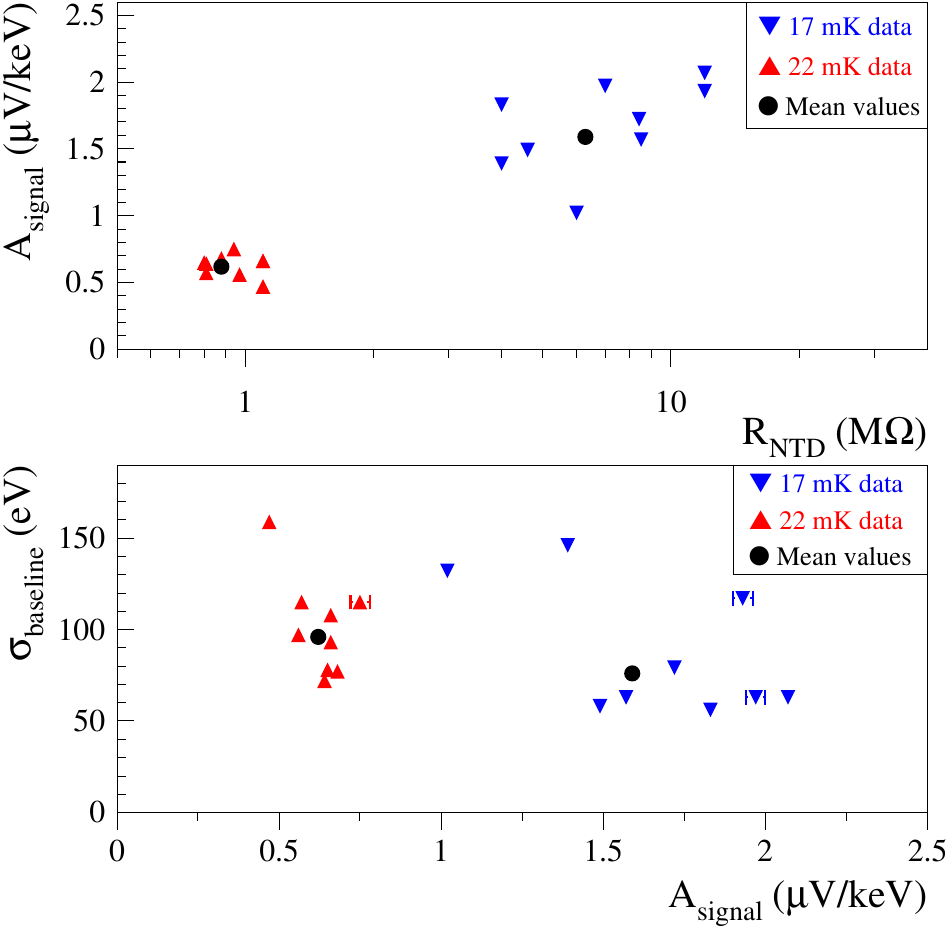}
\caption{Resistance-dependent sensitivity of Ge light detectors (top panel) and its impact on the baseline energy resolution (bottom panel) measured at 17 and 22 mK. A strong NTD polarization was also applied for the light detectors operation at the warmer temperature. 
Statistical uncertainties of the quoted values are (mostly) hidden by the data points.}
\label{fig:LD_performance}
\end{figure}  

At the ``warmer'' temperature and with a significantly higher NTD current (10 nA), the working resistances of the thermistors were reduced to $\sim$1 M$\Omega$, a factor 5--10 lower compared to the ``colder'' operation. At the same time, the sensitivity of LDs is decreased by a factor 2.5, to 0.47--0.75 $\mu$V/keV (0.62 $\mu$V/keV mean). 
Despite a drastic reduction in sensitivity, the energy-converted noise is found to be worse by only ten(s) \% for most of the devices, in the range 0.07--0.16 keV RMS (0.09 keV RMS mean), as illustrated in figure \ref{fig:LD_performance}. 

\begin{figure}
\centering
\includegraphics[width=0.7\textwidth]{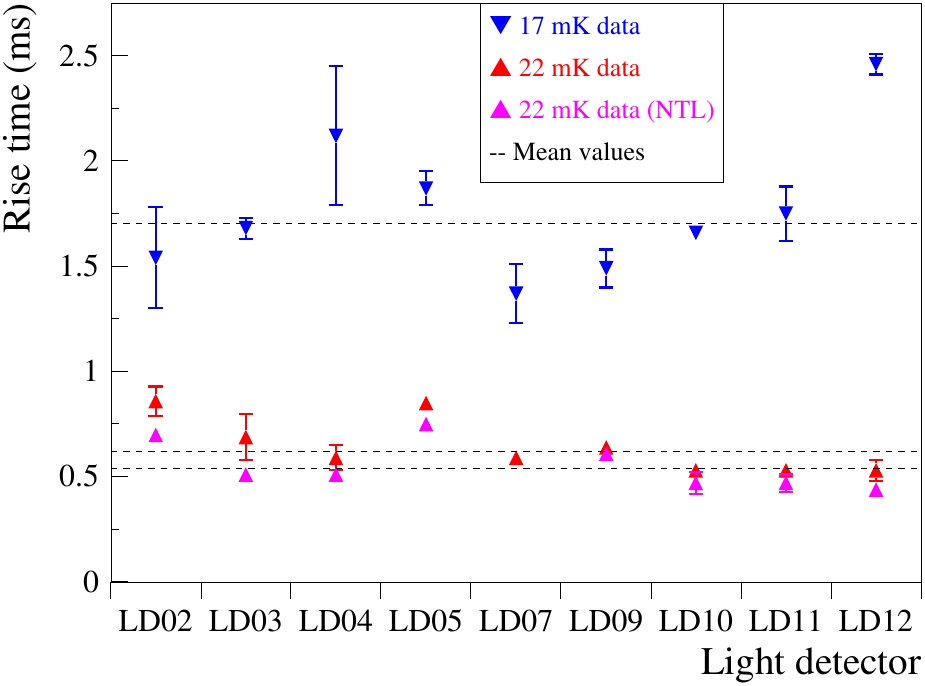}
\caption{Rise-time values of thermal signals of Ge light detectors operated at 17 and 22 mK. Results obtained at 22 mK in the NTL mode of LDs (see section \ref{sec:LD_NTL_studies}) are also shown for comparison.}
\label{fig:LD_risetime}
\end{figure}

In order to characterize the LD thermal response, we use two pulse-shape parameters, rise time ($\tau_{rise}$) and decay time ($\tau_{decay}$), of LED-induced events with tens keV energy, having the same pulse shape as scintillation/particle interactions in LDs. 
As a result of the warmer working temperatures of NTDs, the rising part of thermal signals (i.e., $\tau_{rise}$), is in the range 0.5--0.9 ms (0.7 ms mean), that is more than twice shorter than those of the colder measurements, while the descending part (i.e. $\tau_{decay}$) is in the range 2--6 ms and remains less affected (see figure \ref{fig:LD_risetime}). It is worth emphasizing the sub-millisecond rise times of LDs achieved in the present work: such a fast response is of special importance in view of pile-up rejection (as discussed in section \ref{sec:Prospects}). 
At the same time, dedicated measurements of the scintillation time constants of Li$_2$MoO$_4$ crystals \cite{Spassky:2015,Casali:2019,Mailyan:2023,Bratrud:2025} indicate that these rise times are still an order of magnitude above the dominant photon emission time-scales in Li$_2$MoO$_4$s and allow for further optimization of the detector time response.

%----------------------------------------------------------------
\subsection{Detection of scintillation light}
\label{sec:LY_calibration}

Detection of scintillation light emitted by the coupled crystals can be exploited for the evaluation of the performance of LDs operated in the NTL amplification mode (described in the next section). To find coincidences on the LD-data stream, we used the time positions of events triggered by massive bolometers and took into account an order of magnitude faster rise times of the LDs, similar to the method described in \cite{Piperno:2011}. The scintillation signals' amplitudes found are calibrated using the conversion parameters established with the X-ray measurements (see an example in figure \ref{fig:LD_calibration}). We then construct a dimensionless parameter, Light Yield ($LY$), defined as the ratio of the LD signal to the heat signal (typically expressed in keV/MeV). This parameter represents the proportion of energy carried by the scintillation light detected by the LD relative to the heat energy released in the Li$_2$MoO$_4$ crystal. This ratio varies according to the type of particle. 

\begin{figure}
\centering
\includegraphics[width=0.95\textwidth]{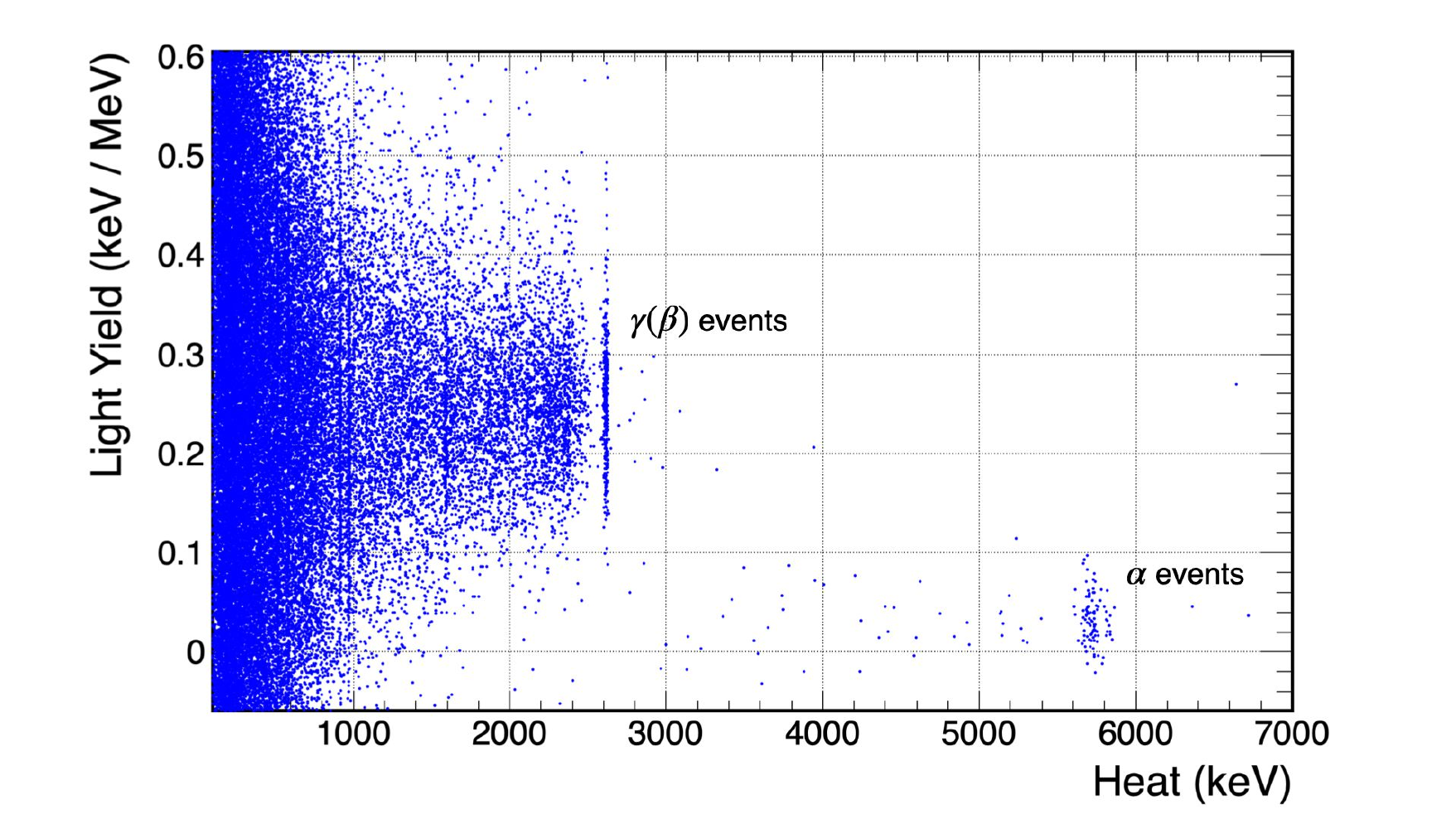}
\caption{Distribution of the $LY$ parameter as a function of the Heat energy of particles detected by the reference Li$_2$$^{100}$MoO$_4$ (Li$_2$MoO$_4$-76b) bolometer in coincidences with a ``close'' Ge light detector (LD-11). Data were collected at 17 mK over 83 h of $^{232}$Th calibration measurements. The energy scale of the Li$_2$$^{100}$MoO$_4$ bolometer was calibrated with $\gamma$ quanta, thus $\alpha$ particles (mainly originated to detector bulk/surface contamination by $^{210}$Po), seen in the 3--6 MeV region with $LY$ below 0.1 keV/MeV, have around 8\% miscalibration (thermal quenching) typical for such detector material \cite{Armengaud:2017}.}
\label{fig:Li$_2$MoO$_4$_LY_Heat}
\end{figure}  

Figure \ref{fig:Li$_2$MoO$_4$_LY_Heat} provides an example of a typical energy-dependent distribution of the $LY$ parameter for Li$_2$MoO$_4$-based scintillating bolometers, operated without a reflective cavity (i.e., an open detector structure) and before applying an NTL bias to LDs electrodes. The majority of events present in this figure belongs to $\gamma$ or $\beta$ interactions in the crystal scintillator, forming a band with a mean $LY$ value of 0.25 keV/MeV. 
Similar results are reported for this type of material tested in similar light collection conditions (including the previous tests of the reference detectors \cite{CROSSdetectorStructure:2024}). 
A population of high-energy events (e.g., above $\sim$3 MeV) with a mean $LY$ value of 0.05 keV/MeV, i.e., only 20\% of light produced for $\gamma$($\beta$)'s of the same energy, corresponds to $\alpha$ decays (mainly $^{210}$Po) in the crystal and on surfaces of the close detector components. Despite a modest light collection efficiency (a factor 3 lower $LY$ than that with a reflective film \cite{Poda:2021}) and a comparatively high noise of the coupled LD (about 110 eV RMS), figure \ref{fig:Li$_2$MoO$_4$_LY_Heat} shows a satisfactory capability of particle identification, providing a high enough efficiency to discriminate $\alpha$'s from $\gamma$($\beta$) events. These two populations can be further separated thanks to the improved LDs' performance achievable with the NTL amplification, which is detailed in the next section.

\begin{figure}
\centering
\includegraphics[width=0.7\textwidth]{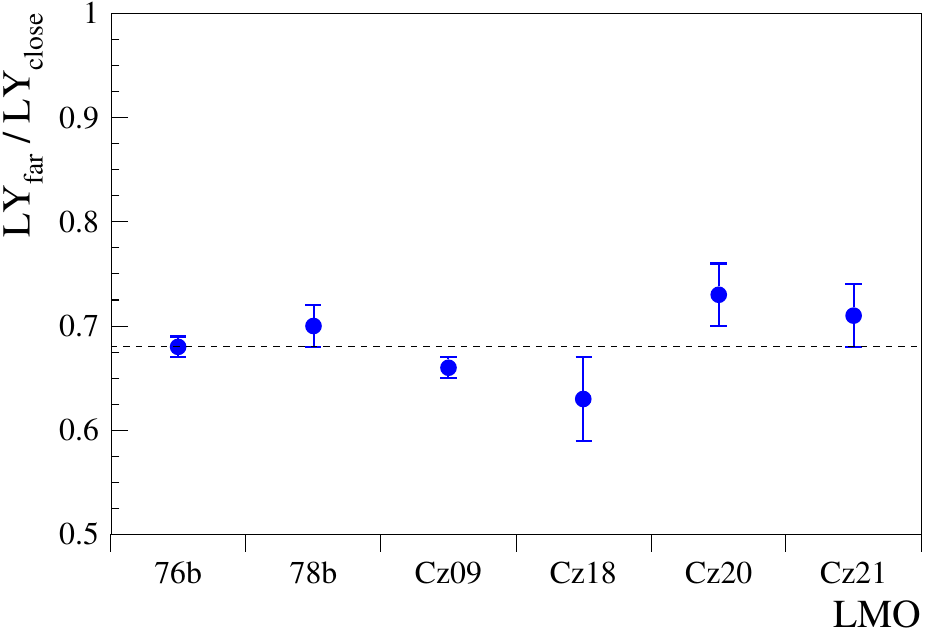}
\caption{Ratio of the $LY_{\gamma(\beta)}$ parameters ($LY_{far}$ / $LY_{close}$) for each Li$_2$MoO$_4$-based scintillating bolometer characterized in the present work. A harmonic mean value (0.68) is shown by a dashed line.}
\label{fig:Li$_2$MoO$_4$_LY_far2close}
\end{figure}

Using heat-scintillation coincidences, we extracted mean $LY$ values of $\gamma$($\beta$) events detected by other Li$_2$MoO$_4$-based scintillating bolometers of the 10-crystal tower, as reported in table \ref{tab:Li$_2$MoO$_4$_LY}. The light signal (dominated by the Cherenkov radiation) from the tested $^{130}$TeO$_2$ crystals is expected to be a factor 15 lower than that of the scintillation of the reference Li$_2$$^{100}$MoO$_4$ crystals \cite{Poda:2021} (i.e., 0.02 keV/MeV for a ``close'' LD), and it was detected only in the NTL mode of LDs \cite{CROSS_enriched_TeO:2024}. For each investigated Li$_2$MoO$_4$-based sample, we present the $LY_{\gamma(\beta)}$ estimates from the top (``close'') and bottom (``far'') Ge LDs, with 0.8 and 6.5 mm distances from the corresponding surface of the coupled crystal. The scintillation detected by the ``close'' LDs of the reference crystals is similar to the results reported for earlier this detector structure \cite{CROSSdetectorStructure:2024}. The increased distance, together with the shadowing of the Cu frame, reduces the scintillation signal detected by the ``far'' LD by about 30\%. Natural samples, provided by a different supplier than the one producing the enriched crystals, exhibit lower scintillation efficiency compared to the reference samples, while maintaining a similar ratio (2/3) between the scintillation signals of the ``far'' and ``close'' LDs, as illustrated in figure \ref{fig:Li$_2$MoO$_4$_LY_far2close}.

\begin{table}
\centering
\caption{Light Yield values of $\gamma$($\beta$) events detected by Li$_2$MoO$_4$-based scintillating bolometers operated in the 10-crystal tower of the CROSS detector assembly design. For each crystal we report two values of the $LY_{\gamma(\beta)}$ parameter depending on the distance of Ge light detectors facing a crystal scintillator:  $LY_{close}$ and $LY_{far}$ measured by the ``close'' (top) and ``far'' (bottom) LD respectively (see details in text). Uncertainties of the $LY$ values are taken (conservatively) as a width (RMS) of their distributions. The last column reports the ratio of the $LY_{\gamma(\beta)}$ values (statistical uncertainties are quoted).}
\smallskip

\begin{tabular}{cccccc}
\hline
Detector  & $LY_{close}$ & $LY_{far}$ & $LY_{far}$ / $LY_{close}$  \\
ID & (keV/MeV) & (keV/MeV) & ~  \\
\hline
\hline
Li$_2$MoO$_4$-76b    & 0.25(6) & 0.17(3) & 0.68(1)    \\ % m703
%\hline
Li$_2$MoO$_4$-78b    & 0.27(4) & 0.19(7) & 0.70(2)  \\ % m703
\hline
Li$_2$MoO$_4$-Cz09   & 0.21(3) & 0.13(3) & 0.66(1)    \\ % m703
%\hline
Li$_2$MoO$_4$-Cz18   & 0.14(5) & 0.09(3) & 0.63(4)    \\ % m703
%\hline
Li$_2$MoO$_4$-Cz20   & 0.14(3) & 0.10(4) & 0.73(3)    \\ % m703
%\hline
Li$_2$MoO$_4$-Cz21   & 0.15(5) & 0.11(3) & 0.71(3)    \\ % m703
\hline
\hline
Mean       & ~      & ~      & 0.68(2)    \\ % m703
\hline
\end{tabular}
\label{tab:Li$_2$MoO$_4$_LY}
\end{table}

%----------------------------------------------------------------
\subsection{NTL amplification for light detectors}
\label{sec:LD_NTL_studies}

\begin{figure}
\centering
\includegraphics[width=1.0\textwidth]{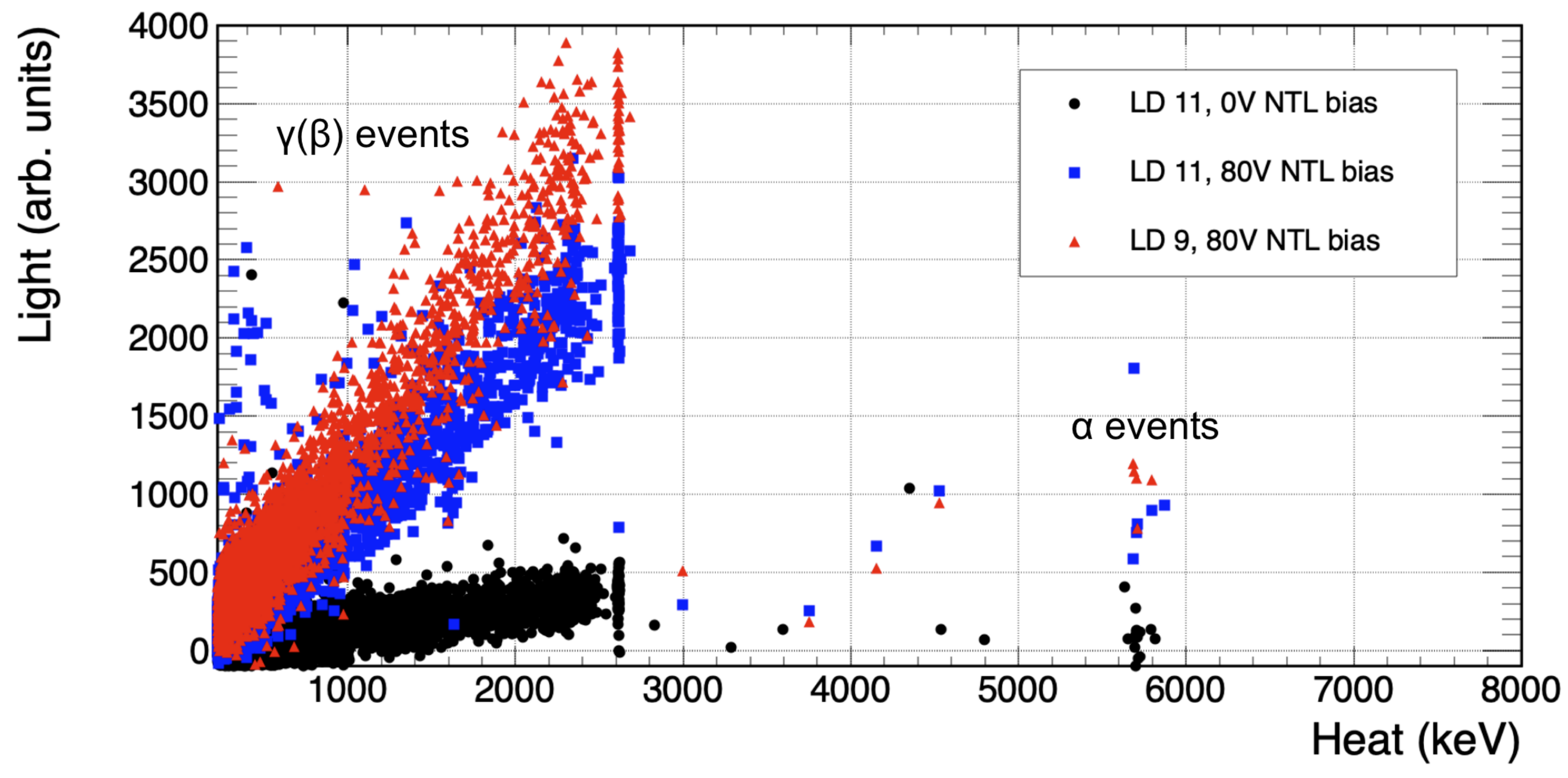}
\caption{Particle identification achieved with the top (LD-11) and bottom (LD-9) LDs, operated in the NTL mode (80 V electrode bias), in coincidences with the events measured by the reference Li$_2$$^{100}$MoO$_4$ scintillating bolometer (Li$_2$MoO$_4$-76b). The absence of the NTL amplification (0 V electrode bias) is illustrated by the top LD, while the bottom LD detected about 30\% lower signal because of the design limitation (see details in text). The 80~V (0~V) data were collected at 22 mK over 18~h (19~h) of $^{232}$Th calibration measurements. The energy scale of the Li$_2$$^{100}$MoO$_4$ bolometer was calibrated with $\gamma$ quanta, thus $\alpha$ particles, seen in 3--6 MeV region, are characterized by 8\% miscalibration.}
\label{fig:Li$_2$MoO$_4$_LD_NTLLD}
\end{figure}

While characterizing the NTL mode of bolometric Ge LDs, we found that all but one operational LD can withstand an electrode bias of $V_{NTL}$ = 100 V without developing significant leakage current (whereas the exception, LD-7, exhibits high leakage at $V_{NTL}$ $>$ 50 V). However, the optimal SNR for most devices was achieved at a slightly lower electrode bias, $V_{NTL}$ = 80 V, consistent with early NTL-LDs studies \cite{Novati:2019}. This voltage was subsequently applied in parallel to eight LDs (with LD-6 and LD-7 grounded). Given that X-ray calibration is no longer possible in the NTL mode of LDs \cite{Novati:2019}, we calibrated them based on their previously established detected scintillation light yield at $V_{NTL}$ = 0 V 
(i.e., using the $LY$ values described in section \ref{sec:LY_calibration} and detected under identical light collection conditions). The results of the NTL LDs' characterization are presented in table \ref{tab:NTLLD_performance} and illustrated in figure \ref{fig:Li$_2$MoO$_4$_LD_NTLLD} for a single module.

\begin{table}
\centering
\caption{Performance of bolometric Ge light detectors with signal amplification based on the Neganov-Trofimov-Luke effect (22~mK data, 80 V electrode bias). The LD calibration is performed by scintillation light of ``close'' and ``far'' Li$_2$MoO$_4$s (data of  table \ref{tab:Li$_2$MoO$_4$_LY}). The effective NTL gain ($G_{NTL}^{eff}$) is given with respect to the initial LDs performance at 22~mK and no NTL effect (see table \ref{tab:LD_performance}). The width of $\tau_{rise}$ distribution is taken as an uncertainty of the LD response, while statistical uncertainties of the $A_{signal}$, $\sigma_{baseline}$, and $G_{NTL}^{eff}$ values are quoted.}
\smallskip
\begin{tabular}{lccccccc}
\hline
Detector  & $\tau_{rise}$ & \multicolumn{2}{c}{$A_{signal}$ ($\mu$V/keV)} & \multicolumn{2}{c}{$\sigma_{baseline}$ (eV)} & \multicolumn{2}{c}{$G_{NTL}^{eff}$}  \\
ID & (ms) & Close & Far & Close & Far  & Close & Far   \\
\hline
\hline
LD-2    & 0.70(2) & 3.37(2) & 5.40(4) & 17.0(2) & 10.6(1) & 4.59(3) & 7.36(5)   \\ % m813
%\hline
LD-3    & 0.51(2) & 2.03(2) & -- & 32.4(2) & -- & 3.55(3) & --   \\ % m813
%\hline
LD-4    & 0.51(2) & 2.65(3) & 8.16(6) & 25.1(2) &  8.1(1) & 3.86(5) & 12.0(1)   \\ % m813
%\hline
LD-5    & 0.75(1) & 3.37(11) & 8.77(13) & 16.8(2) &  6.4(1) & 4.30(14) & 11.3(2)   \\ % m813
%\hline
LD-9    & 0.61(2) & 3.77(9) & 6.80(10) & 16.3(1) &  9.0(1) & 4.72(11) & 8.56(12)   \\ % m813
%\hline
LD-10   & 0.47(5) & 2.77(4) & 6.67(4) & 32.5(3) & 13.5(1) & 2.86(5) & 6.89(4)   \\ % m813
%\hline
LD-11   & 0.47(4) & 3.24(4) & 6.00(4) & 24.5(3) & 13.2(1) & 4.41(5) & 8.18(6)   \\ % m813
%\hline
LD-12   & 0.44(1) & 2.45(2) & -- & 39.6(5) & -- & 4.02(3) & --   \\ % m813
\hline
\hline
Mean    & 0.54(2) & 2.85(3)  & 6.78(6) & 23.0(2) & 9.5(1) & 3.94(5)  & 8.66(7)   \\ % m813
\hline
\end{tabular}
\label{tab:NTLLD_performance}
\end{table}

\begin{figure}
\centering
\includegraphics[width=0.7\textwidth]{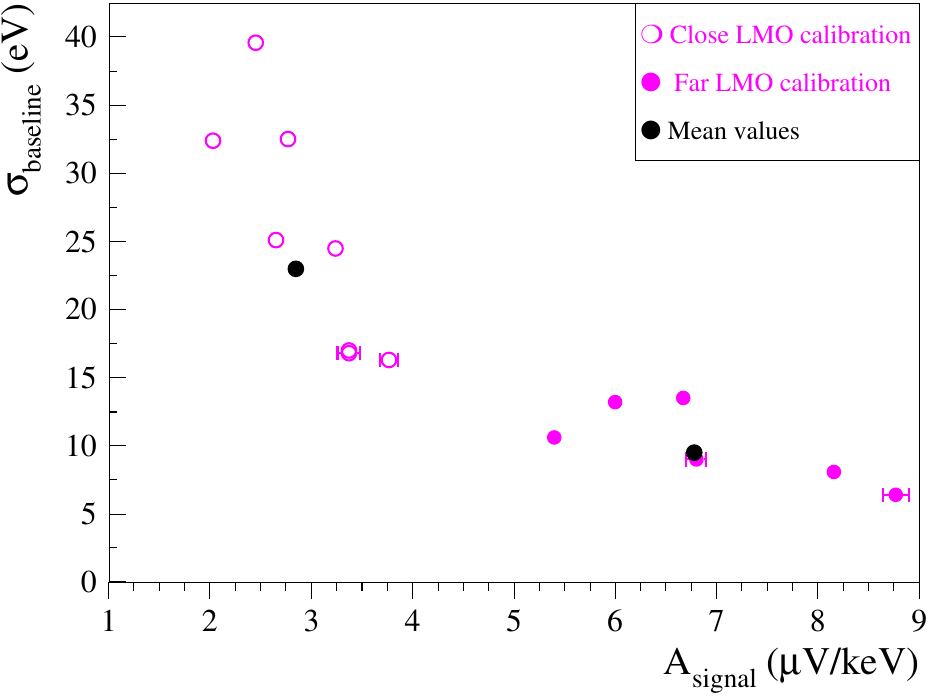}
\caption{Baseline noise resolution (RMS) versus sensitivity of Ge light detectors, operated in the Neganov-Trofimov-Luke amplification mode (80 V electrode bias) at 22 mK. The energy scale was determined using the $LY_{\gamma(\beta)}$ values of ``close'' and ``far'' Li$_2$MoO$_4$ crystal. The two mean values correspond to ``close'' (left black point) and ``far'' (right black point) configurations.}
\label{fig:NTLLD_performance}
\end{figure}

\begin{figure}
\centering
\includegraphics[width=0.7\textwidth]{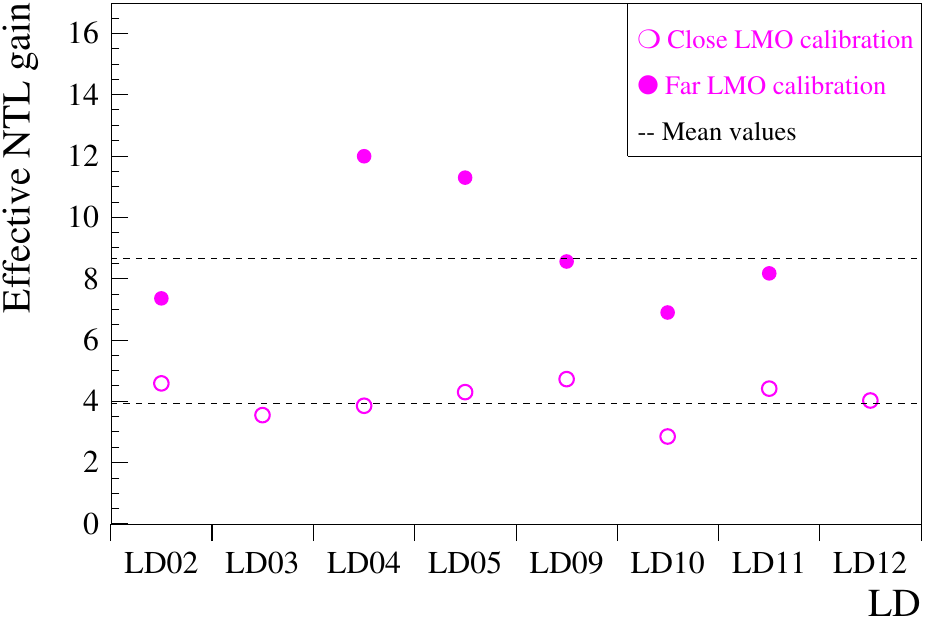}
\caption{Effective Neganov-Trofimov-Luke gain of thermal signals of bolometric Ge light detectors (22 mK data, 80 V NTL bias), with respect to their position (``close'' or ``far'') to the neighbor Li$_2$MoO$_4$ scintillation crystal.}
\label{fig:NTLLD_gain}
\end{figure}

After applying the NTL bias (80 V), we observed that the LDs' response (represented by LED-induced signals) became slightly faster, which can be explained by an impact of the SNR on the computation of the parameter in combination with non-linearity of the devices. We obtained $\tau_{rise}$ values in the range 0.42--0.74 ms, with a mean of 0.54(2) ms (for example, the mean of 0.62(3) ms is reported above for the 0 V case). 

Thanks to the NTL amplification, the sensitivity of LDs increases by a factor $G_{NTL}$ = (4--10) compared to the initial performance of the devices (listed in table \ref{tab:LD_performance}). 
We remark that the gain measured with the scintillation light is lower than that one could obtain with a full coverage of the wafer with Al electrodes. In the present configuration, only $\sim$56\% of the wafer's surface with electrodes is involved in the NTL amplification. We therefore expect, under the same applied voltage, an improvement of the light signal amplitude by an area factor approximately 1.7 (see section \ref{sec:Simulations}) with an appropriate electrode configuration covering the whole surface. 
Also, it is interesting to note that despite a 30\% lower light collection for ``far'' LDs, the NTL amplification of their signals is approximately twice as high as that of ``close'' LDs, which detect scintillation by the Ge side having no Al electrodes. 
This shows that 
with the current electrode configuration, NTL is a surface phenomenon. Its efficiency is dependent on the field lines and by the trapping centers for electrons and holes due to surface defects.

As a result of the improved sensitivity, a substantial noise reduction of NTL-LDs ($\sigma_{baseline}$) was achieved, allowing a mean value of 24 eV and 9.5 eV RMS for the calibration with ``close'' and ``far'' Li$_2$MoO$_4$s, respectively. A correlation between LD-NTL sensitivity and noise is clearly seen in figure \ref{fig:NTLLD_performance}.  
The effective NTL gain $G_{NTL}$ is defined as the improvement in SNR. It remains similar to the gain in LD sensitivity as listed in table \ref{tab:NTLLD_performance} and illustrated in figure \ref{fig:NTLLD_gain}. 
Therefore, the NTL effect does not induce substantial excess noise in the NTL voltage range adopted here.

%==============================================================================
\section{Prospects for large-scale $0\nu\beta\beta$ search experiments}
\label{sec:Prospects}

%----------------------------------------------------------------
\subsection{CROSS detector structure and NTL-LD technology}

The present and early \cite{CROSSdetectorStructure:2024} studies of Li$_2$MoO$_4$-based scintillating bolometers assembled in the CROSS mechanical structure clearly demonstrate the feasibility of high detector performance in an array containing a low amount of passive materials 
in compliance with the demands of CROSS, CUPID and other possible future large-scale bolometric $0\nu\beta\beta$ experiments \cite{Alfonso:2022, Alfonso:2025gdpt}. 

\begin{figure}
\centering
\includegraphics[width=0.95\textwidth]{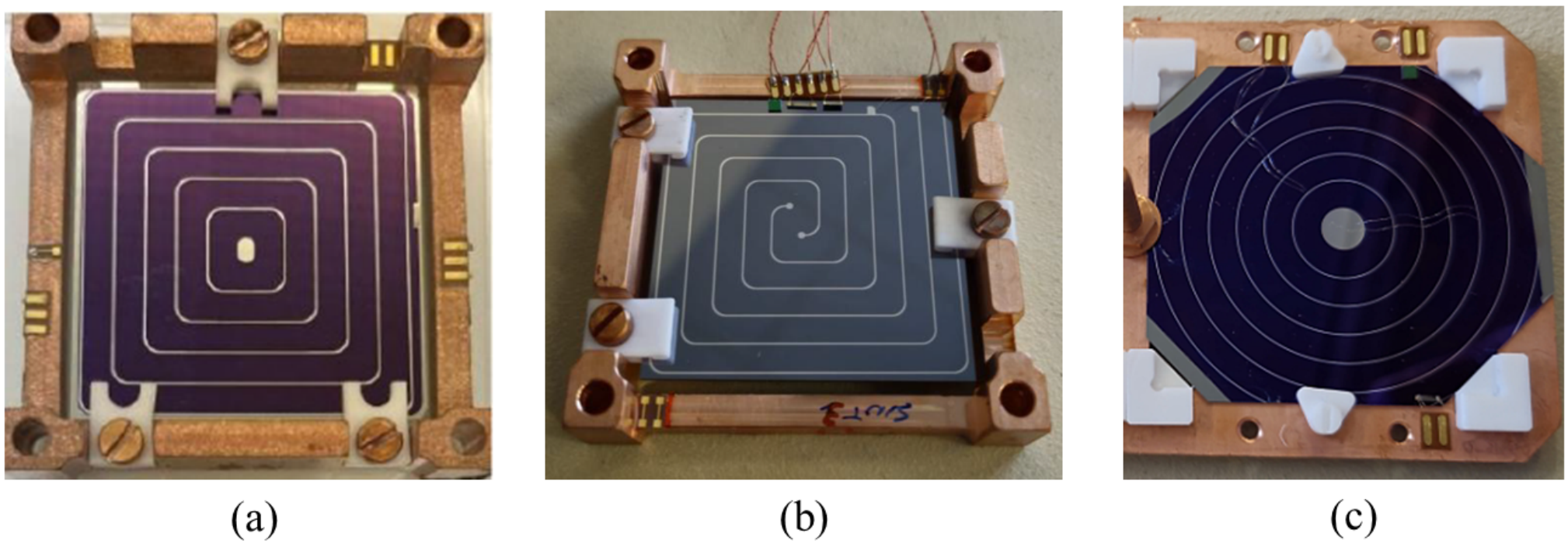}
\caption{Prototypes of NTL-LDs to be used in the CROSS demonstrator and in BINGO, based on square-shaped wafers with Al electrodes of semi-square (a) and dual-spiral (b) geometry enabling a full coverage of wafer's surface. A prototype of the CUPID Baseline NTL-LD of octagonal shape together with circular electrodes (c), which will be produced for the second CUPID prototype tower.}
\label{fig:NTLLD_CROSS_CUPID}
\end{figure}

Also, the above-described test of NTL LD devices proved that the proposed technology is compatible with a large array and it will be transferred to the fabrication of such devices for CUPID \cite{CUPID_NTLLD:2024}. 
This knowledge transfer is expected to be straightforward, taking into account that the same absorber material with similar size will be used in the fabrication of CUPID LDs \cite{Alfonso:2023}. Moreover, the circular electrode design combined with the slightly larger area of octagonal-shaped Ge LDs of the CUPID Baseline design would enable the use of an additional circular Al electrode, increasing the area for the NTL effect from 56\% (for a square-shaped wafer) to 75\%. An R\&D on the electrode design is ongoing to extend the NTL electrode coverage to 100\% \cite{CUPID_NTLLD:2024}. These advances in electrode design and wafer size are illustrated in figure \ref{fig:NTLLD_CROSS_CUPID}. The present (circular) electrode design, together with the new ones (semi-square and dual-spiral), will be used in the 42-crystal CROSS demonstrator, while larger, octagonal, LDs with circular electrodes will be used in the second CUPID prototype tower with 28 crystals. In the following sections, we consider a possible impact of the NTL-LD technology on the CUPID objectives in terms of background components. Of course, these considerations are relevant for any $0\nu\beta\beta$ experiment based on scintillating bolometers containing molybdenum and with a relatively slow LD readout as that provided by NTD Ge thermistors.

%----------------------------------------------------------------
\subsection{Background reduction with NTL-LDs} 

Thanks to the NTL amplification, the $>$99.9\% rejection of the alpha particles at the $^{100}$Mo $0\nu\beta\beta$ ROI is well-secured. The amplification provides also an additional robustness with respect to higher noise level. 
A fast time response combined with the enhanced SNR of LDs can provide a viable tool 
for rejection of another dominant background in CUPID and Mo-based bolometric experiments -- random coincidences of $2\nu\beta\beta$ events from decays of $^{100}$Mo inside the crystals \cite{Chernyak:2012, Chernyak:2014, Chernyak:2017, CROSSpileup:2023} -- as discussed in this section.

Given the results of the NTL-LD investigation in the present work (table \ref{tab:NTLLD_performance}), we can extrapolate the impact of the performance achieved on the CUPID background objectives \cite{Alfonso:2025cupid, CUPID_Bkg_Model:2024}. According to \cite{Chernyak:2012}, the background rate induced by $2\nu\beta\beta$ random coincidences can be calculated based on the pile-up rejection capability. The latter parameter can be determined in turn as a function of the SNR and $\tau_{rise}$. Recently, a study has been conducted \cite{CROSSpileup:2023} utilizing a data streaming from an actual NTL-LD, operated in the same setup as current devices, and generated synthetic pulses on top of baseline traces, modifying the pulse amplitude and shape. A similar analysis was conducted with pulses induced in Li$_2$$^{100}$MoO$_4$ bolometers by heaters glued to the crystal surface \cite{Armatol:2021}. Various PSD parameters were developed and tested to assess their ability to reject adjacent pulses as a function of pulse amplitudes and time separation, keeping the acceptance for single pulses higher than 90\%. The authors of \cite{CROSSpileup:2023} found that the $2\nu\beta\beta$ pile-up rate in the ROI can be reduced below 0.5 $\times$ $10^{-5}$ ckky (CUPID goal \cite{Alfonso:2025cupid}) by applying a PSD to the NTL-LD channel, characterized by $\tau_{rise}$ = 0.8 ms and SNR = 140. The values of SNR are computed in this context considering the ratio between a light signal amplitude for an event in the ROI and the RMS baseline width, after application of the optimum filter \cite{Gatti:1986}. 

As seen in table \ref{tab:NTLLD_performance}, the results of the NTL-LDs in terms of rise times are satisfactory: most of the operated devices demonstrated $\tau_{rise}$ = (0.4--0.5) ms, a factor 2 faster than the best results reported in \cite{CROSSpileup:2023} and used there for simulations. The high bias currents mentioned above are mandatory to obtain this fast response. 
Considering to flip the LD with the Al electrode side facing the closer crystal we should expect a total SNR at $Q_{\beta\beta}$ of $^{100}$Mo (3034 keV) corresponding to $A_{light}$/$\sigma_{baseline}$  = 0.3$\cdot$3.034/0.01 $\sim$ 90, using the $LY_{\gamma(\beta)}$ = 0.3 keV/MeV of the ``close'' geometry \cite{CROSSdetectorStructure:2024} and the baseline resolution of 10 eV (a mean achieved in the present study). 
According to \cite{CROSSpileup:2023}, this combination of the $\tau_{rise}$ and SNR values could potentially comply with the CUPID background goal. However, these pile-up rejection studies \cite{CROSSpileup:2023} were performed with another NTL-LD (smaller sizes of the Ge wafer and NTD Ge thermistor, independent holder) and under noise conditions which are not fully comparable to the present measurements (in particular, the present data are characterized by a higher contribution of high-frequency noise which affects the bandwidth of LDs). Therefore, we decided to perform an independent investigation of the pile-up-induced background with the noise data from all 8 devices analyzed in this work, following the approach of Ref. \cite{CROSSpileup:2023} and considering several configurations for CUPID, as detailed in the next section. 

%----------------------------------------------------------------
\subsection{Simulations of pile-up induced background} 
\label{sec:Simulations}

For simulations of pile-up-induced background in view of CUPID we considered several configurations of NTL-LDs and detector structures which are listed in table \ref{tab:NTLLD_config_CUPID}. The configuration (I) corresponds to the detector structure investigated in the present work, and its upgrade to NTL-LDs with larger surface coverage by an electrode is given by (II). We observe that the configuration (II) closely resembles the one that will ultimately be adopted by the BINGO experiment~\cite{Armatol:2024bingo}. The last two configurations, (III) and (IV), respectively, represent versions of the CUPID Baseline structure containing NTL-LDs with circular electrodes ($\sim$75\% of coverage) and with full surface coverage. The light collection efficiency has been taken for the closest NTL-LD, which corresponds to a mean $LY_{\gamma(\beta)}$ of 0.30 and 0.36 keV/MeV for the CROSS \cite{CROSSdetectorStructure:2024} and CUPID Baseline \cite{Alfonso:2025gdpt} structures, respectively. It is also assumed that the electrodes of NTL-LDs are facing crystals (to maximize the NTL gain). 

\begin{table}
\centering
\caption{Configurations of detector structure and electrode coverage considered for pile-up simulations.}
\smallskip
\begin{tabular}{cccccc}
\hline
Configuration & Detector & LD & NTL electrode & $LY_{\gamma(\beta)}$   & Scaling factor \\
~ & structure & shape & coverage  & (keV/MeV)  &  $f_{scale}$ \\
\hline
\hline
(I)    & CROSS             & square    &  56\%  & 0.30 & 1.0 \\
(II)   & CROSS             & square    & 100\%  & 0.30 & 1.7 \\
(III)  & CUPID Baseline    & octagonal &  75\%  & 0.36 & 1.6 \\
(IV)   & CUPID Baseline    & octagonal & 100\%  & 0.36 & 2.0 \\
\hline
\end{tabular}
\label{tab:NTLLD_config_CUPID}
\end{table}

To account for the impact of different electrode coverages on the performance of NTL-LDs (i.e., a gain of SNR), we defined a scaling factor $f_{scale}$ of the NTL gain with respect to the configuration (I) investigated experimentally. The impact of the partial electrode coverage on the observed NTL gain in the configuration (I) can be expressed as:
\begin{equation}
    G_{CROSS} = (1 - \xi_{CROSS}) + \xi_{CROSS} \cdot G_{NTL},
\label{eq:Gcross}
\end{equation}
where $G_{CROSS}$ is the effective NTL gain achieved in the present study with NTL-LDs in the CROSS structure (with a mean of 8.66 for NTL-LDs with electrodes facing the scintillation source), $\xi_{CROSS}$ is the fraction of the area of a square-shaped LD covered by circular Al electrodes ($\xi_{CROSS}$ = 0.56), and $G_{NTL}$ is the NTL gain in the electrode covered area. Thus, the scaling factor for the configuration (II) is defined by the following equation:
\begin{equation}
    f_{scale} = \frac{G_{NTL}}{G_{CROSS}} = \frac{1}{\xi_{CROSS}} - \frac{1 - \xi_{CROSS}}{\xi_{CROSS} \cdot G_{CROSS}},
\label{eq:fcross}
\end{equation}
and is calculated as $f_{scale}$ = 1.7. 
Similarly to Eq. \ref{eq:Gcross}, we can represent a possible NTL gain for the devices of the CUPID Baseline design, which can be produced adopting the above described technology:
\begin{equation}
    G_{CUPID} = (1 - \xi_{CUPID}) + \xi_{CUPID} \cdot G_{NTL},
\label{eq:Gcupid}
\end{equation}
where $G_{CUPID}$ is the effective NTL gain of LDs in the CUPID structure, $\xi_{CUPID}$ is a fraction of the electrode covered area for octagonal LDs. So, the scaling factor for these configurations can be calculated as follows:  
\begin{equation}
    f_{scale} = \frac{G_{CUPID}}{G_{CROSS}} \cdot \frac{LY_{CUPID}}{LY_{CROSS}} ,
\label{eq:fcupid_0}
\end{equation}
\noindent where $LY_{CUPID}$ and $LY_{CROSS}$ are average $LY_{\gamma(\beta)}$ values for these detector structures (0.36 and 0.3 keV/MeV, respectively).
Using Eqs. \ref{eq:Gcross} and \ref{eq:Gcupid}, the scaling factor for CUPID configurations can be expressed as
\begin{equation}
    f_{scale} = \left(\frac{1 - \xi_{CUPID}}{G_{CROSS}} + \xi_{CUPID} \cdot \left(\frac{1}{\xi_{CROSS}} - \frac{1 - \xi_{CROSS}}{\xi_{CROSS} \cdot G_{CROSS}}\right)\right) \cdot \frac{LY_{CUPID}}{LY_{CROSS}}.
\label{eq:fcupid}
\end{equation}
Thus, taking $\xi_{CUPID}$ equal to 0.75 and 1.0 for configurations (III) and (IV), one can obtain $f_{scale}$ = 1.6 and 2.0, respectively.

A pile-up simulation was performed following the procedure described in \cite{CROSSpileup:2023} with some modifications detailed below. A single high-energy pulse of each NTL-LD was taken as a pulse template for signal injections. A 0.5-s-long window is used for the template, which is much longer than the typical decay time of the pulse (below 10 ms). The energy of the pulses is randomly selected from the $2\nu\beta\beta$ distribution of $^{100}$Mo; the ROI is taken as $E$ = [3019; 3049] keV, that is, $\pm$15 keV around  $^{100}$Mo $Q_{\beta\beta}$. The corresponding signal amplitude in the light channel is defined using the $LY_{\gamma(\beta)}$ value (for the CROSS structure), the scaling factor $f_{scale}$, and the individual performance of the considered NTL-LD; the number of created photons (with energy of 2.1 eV, characteristic of the Li$_2$MoO$_4$ emission) is extracted from a Poisson distribution to account for the fluctuation of the number of produced photons. We performed simulations for all configurations except (I), which is characterized by a factor 2 lower $f_{scale}$ than the others. 
Instead of generating $N_{inj}$ pile-ups for each $\Delta$t, as done in \cite{CROSSpileup:2023}, we inject $N_{inj}$  = 16000 for $\Delta$t uniformly distributed between [0; $\Delta$t$_{r = 100\%}$], where $\Delta$t$_{r = 100\%} \approx \tau_{rise}$, corresponding to a rejection power $r$ = 100\% (this assumption was tested).  
Then, we process the generated pile-up data, apply PSD cuts and define the rejection power $r$ as a ratio of a number of rejected pile-ups $N_{rej}$ to the total number of injected events: 
\begin{equation}
    r = \frac{N_{rej}}{N_{inj}}.
\label{eq:rejection}
\end{equation}
We checked that $r$ does not depend on the choice of $\Delta$t$_{r = 100\%}$ as far as the rejection is actually 100\% for $\Delta$t$_{r = 100\%}$.

Finally, we calculate the background index $BI$ using the following formula 
\begin{equation}
    BI = P \cdot (1 - r) \cdot \frac{A^2}{m} \cdot Time \cdot \Delta t_{r = 100\%},
\label{eq:BI}
\end{equation}
where $P$ is a probability value to have a pile-up event in the ROI corresponding to the worst case in terms of pile-ups rejection, i.e., $P$ = $P_{\Delta t = 0}$ = 3.41 $\times$ 10$^{-4}$ keV$^{-1}$ \cite{CROSSpileup:2023}, $A$ is the activity of $^{100}$Mo in the considered Li$_2$MoO$_4$ crystal (2.93 mBq), and $Time$ is a 1-yr-long period given in seconds.

The method we used for the pile-up injection is time efficient and the uncertainties inferred by the post-analysis steps should be less because of the simpler procedure of $BI$ calculation (absence of multiple fits and interpolation) and due to the continuous $\Delta$t distribution (no need for discretizations). In addition, we verified with a couple of channels (LD-9 and LD-10) that both methods give similar values of $BI$ (the observed difference is $\sim$1\%); the enlargement of the $\Delta$t interval also leads to consistent results. 

\begin{table}
\centering
\caption{Projection of performance of NTL-LDs, based on the results of the present study and considering several configurations of CUPID detectors (see table~\ref{tab:NTLLD_config_CUPID}). We quote the SNR for the scintillation signal induced by $^{100}$Mo $0\nu\beta\beta$ decays in a Li$_2$$^{100}$MoO$_4$ bolometer, and achievable background index ($BI$) in a 30-keV-wide ROI centered at $^{100}$Mo $Q_{\beta\beta}$ associated with the residual contribution of pile-ups after the pulse-shape discrimination using NTL-LD signals. Uncertainties reported are statistical only (uncertainties of BI values are below the given precision).}
\smallskip
\begin{tabular}{c|cccc|ccc}
\hline
Detector & \multicolumn{4}{c|}{SNR at $^{100}$Mo $Q_{\beta\beta}$}   & \multicolumn{3}{c}{BI ($10^{-5}$ ckky) at $^{100}$Mo $Q_{\beta\beta}$} \\
ID & (I) & (II)  & (III) & (IV) & (II) & (III)  & (IV) \\
\hline
\hline
LD-2    & 81(1) & 138(1) & 128(1) & 166(1) & 7.9 & 8.3 & 7.6 \\ % ch 15
LD-3    & 60(1) & 102(1) & 94(1)  & 122(1)  & 7.9 & 8.3 & 7.7 \\ % ch 19
LD-4    & 115(1) & 196(2) & 179(2) & 235(3) & 5.7 & 5.9 & 5.3 \\ % ch 13
LD-5    & 154(5) & 261(8) & 239(8) & 313(10) & 6.4 & 6.6 & 5.9 \\ % ch 3
LD-9    & 114(3) & 194(5) & 178(4) & 233(6) & 7.1 & 7.3 & 6.4 \\ % ch 5
LD-10   & 64(1) & 108(2) & 100(2) & 130(2) & 6.2 & 6.3 & 5.6 \\ % ch 11
LD-11   & 69(1) & 118(1) & 109(1) & 142(2) & 6.5 & 6.9 & 5.7 \\ % ch 9
LD-12   & 51(1) & 86(1)  & 79(1)  & 103(1) & 6.5 & 6.9 & 5.8 \\ % ch 17
\hline
\hline
Mean   & 78(1) & 132(1) & 122(1) & 159(2) & 6.7 & 7.0 & 6.1  \\ 
\hline
\end{tabular}
\label{tab:NTLLD_performance_CUPID}
\end{table}

\begin{figure}
\centering
\includegraphics[width=1.0\textwidth]{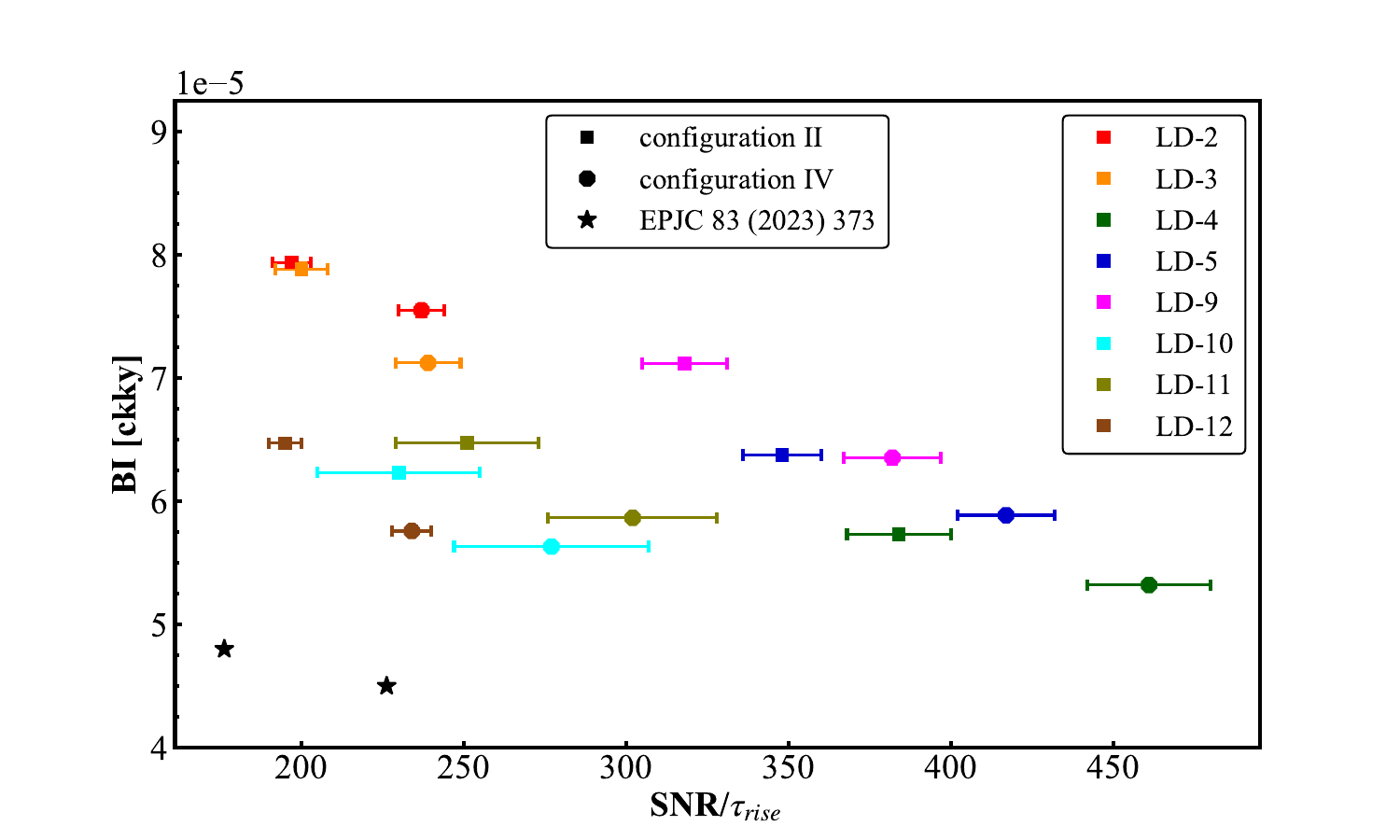}
\caption{Dependence of the background index BI at $^{100}$Mo $Q_{\beta\beta}$ on the ratio of SNR to rise time for the 8 NTL-LDs considered and the two simulated CUPID-like detector configurations (II) and (IV), corresponding respectively to the CROSS and CUPID Baseline detector structures using NTL-LDs with a full electrode coverage of wafers' surface. Results achieved with a similar device operated in the used setup \cite{CROSSpileup:2023} are shown too (stars).}
\label{fig:CUPID_BI_SNR}
\end{figure}  

Table \ref{tab:NTLLD_performance_CUPID} lists the results of the pile-up rejection study, that is a residual background index for different configurations of CUPID detectors. As can be seen, the $BI$ values obtained for the configurations (II) and (III) are very similar, while a further improvement of 10\% is achieved for the settings (IV), resulting in a mean value of 0.6 $\times$ 10$^{-4}$ ckky with several detectors approaching the CUPID background goal.  

Despite these encouraging results, we observe that the background levels estimated through simulations and parameterized by SNR and $\tau_{rise}$ according to \cite{CROSSpileup:2023}, which predicts the $BI \sim 0.5 \times 10^{-4}$ ckky for SNR/$\tau_{rise}$ $\approx$ 180, do not completely account for the specific behavior of individual detectors in terms of noise. This is because the simulations utilized actual noise baseline streaming obtained with a set of typical LDs chosen initially. The noise figure is dominated by vibrations, and a large spread is observed not only in the noise RMS level (captured by SNR) but also in the shape in the frequency domain. In particular, the microphonic peaks and the noise level in the frequency range of 1--20 Hz (and their high-frequency harmonics) change significantly from sample to sample, similar to previous measurements in this setup \cite{CrossCupidTower:2023a}. These observations explain the large spread in the achieved results, as illustrated in figure \ref{fig:CUPID_BI_SNR}, which correlates the LD performance parameters with background levels. It is clear from figure \ref{fig:CUPID_BI_SNR} that there is definitely room for improvement related to noise conditions (e.g. improved grounding of the setup, implementation of noise cancellation techniques) and to the ratio of key parameters SNR/$\tau_{rise}$ (particularly with a higher NTL bias and possibly faster detector response). Both ways of NTL-LD optimization will be investigated in upcoming experiments with the CROSS demonstrator and the second CUPID prototype tower.

Another enhancement in pile-up rejection could be achieved by employing a different transfer function to process signals before applying pulse-shape-parameter cuts. Currently, we utilize a transfer function based on the Gatti-Manfredi optimum filter. However, this signal processing method is optimized to estimate the signal amplitude, and there is no inherent reason to assume it is the best approach for distinguishing pile-up events. In contrast, preliminary tests using a custom-engineered transfer function, designed to emphasize the frequency range where single and overlapping pulses differ most, yielded improved results compared to those presented here. A future publication will explore this new approach in detail once the related work is completed.

%==============================================================================
\section{Conclusions}

Future bolometric experiments to search for $0\nu\beta\beta$ decay (like CROSS, BINGO, and CUPID) will largely rely on high-performance cryogenic light detectors (LDs) for both particle identification and random coincidences rejection. A convenient approach to enhance performance of an ordinary bolometric LD is the deposition of electrodes on its surface to leverage voltage-driven thermal signal amplification based on the Neganov-Trofimov-Luke (NTL) effect. To demonstrate this technology, we fabricated a batch of 10 cryogenic Ge NTL-LDs ($45 \times 45 \times 0.3$~mm each) and 
characterized them in a pulse-tube cryostat at the Canfranc underground laboratory in Spain. The NTL-LDs were assembled in the CROSS-design detector structure together with ten 45-mm side cubic crystals (six Li$_2$MoO$_4$ and four TeO$_2$ based materials). The 10-crystal detector array was operated at 17 mK to optimize detector sensitivity and enable comparison with previous tests in the setup and at 22 mK to optimize time response.

Two $^{100}$Mo-enriched Li$_2$MoO$_4$ bolometers, used as reference thermal detectors of the setup, 
exhibit high bolometric and spectrometric performance (particularly, energy resolution of 1--2 keV FWHM for noise traces and below 6 keV FWHM at 2615 keV), consistent with results from previous tests of these devices in the CROSS setup. 
The noise of the bolometric Ge LDs was measured in a range of 0.05--0.15 keV RMS (0.07 keV mean);  
this performance is compatible with highly efficient scintillation-assisted rejection of $\alpha$-induced background (more than 99.9\%), the dominant contribution to the region of interest for $0\nu\beta\beta$ decay searches with thermal detectors. Operation at a higher temperature (22~mK), combined with relatively high currents across the thermistors, reduced the SNR in the LDs by only $\sim$30\% (despite a 2.5-fold decrease in sensitivity), while the rise time of the thermal signals decreased from $\sim$2 ms (at 17 mK) to the sub-ms range. With an 80 V bias applied to the Al electrodes, the Ge devices achieved a mean rise time of 0.54 ms, and the SNR improved by a mean factor 9 (4 for the surface opposite the electrodes). We note that the gain is limited by the current electrode design (optimized for a circular wafer), which covers approximately 56\% of the area for the NTL effect (i.e., a factor 1.7 compared to the full coverage). Assuming full NTL electrode coverage, the performance of the Ge devices in this work would enable efficient discrimination of pile-ups (primarily induced by two-neutrino double-beta decays of $^{100}$Mo) close to CUPID’s background goals (i.e., a pile-up contribution below $0.5\times10^{-4}$ counts/keV/kg/yr).

%==============================================================================
\acknowledgments

This work was supported by the European Research Council (ERC) under the European Union Horizon 2020 program (H2020/2014-2020) with the ERC Advanced Grant no. 742345 (ERC-2016-ADG, project CROSS) and the ERC Consolidator Grant no. 865844 (Project BINGO). This material is also based upon work supported by the US Department of Energy (DOE), Office of Science under Contract No. DE-AC02-05CH11231 and DE-AC02-06CH11357, and by the DOE Office of Science, Office of Nuclear Physics under Contract No. DE-FG02-00ER41138. 
The authors thank the director and staff of the Laboratorio Subterr\'aneo de Canfranc and the technical staff of our laboratories for providing continuous support. 
Also, the authors are grateful to F.T.~Avignone-III and L.A.~Winslow for coordinating the development of $^{130}$TeO$_2$ and Li$_2$MoO$_4$ crystals, respectively, and to the members of the CUPID collaborations for useful discussions and comments on the manuscript. 
More information on the technical details of the CROSS and BINGO experiments and the respective collaboration policies can be found on the official web pages [\href{https://a2c.ijclab.in2p3.fr/en/a2c-home-en/assd-home-en/assd-cross/}{https://a2c.ijclab.in2p3.fr/en/a2c-home-en/assd-home-en/assd-cross}] and [\href{https://bingo.extra.cea.fr}{https://bingo.extra.cea.fr}], respectively.

%\begin{thebibliography}{99}

%\end{thebibliography}

%\bibliographystyle{plain}
\bibliographystyle{JHEP}
%\bibliography{Bibliography}

\providecommand{\href}[2]{#2}\begingroup\raggedright\endgroup

\end{document}